 \documentclass[nohyper,12pt,letterpaper]{JHEP3}
 \usepackage{epsfig}
 
 \def\hacek{\accent20} 

 \def\fivefivebar{(5,\bar{5})}
 \def\fivebarfive{(\bar{5},5)}
 \title{Embedding the Pentagon}

 \author{T.\,Banks\\
 Department of Physics \\
 University of California, Santa Cruz, CA 95064\\
 E-mail: \email{banks@scipp.ucsc.edu}\\
 {\it and}\\
 NHETC, Rutgers University\\
 Piscataway, NJ 08540}

 \author{S.\,Echols\\
 Department of Physics and SCIPP\\
 University of California, Santa Cruz, CA 95064\\
 E-mail: \email{sechols@physics.ucsc.edu}}

 \author{J. L.\,Jones\\
 Department of Physics and SCIPP\\
 University of California, Santa Cruz, CA 95064\\
 E-mail: \email{jeff@physics.ucsc.edu}}

\abstract{The Pentagon Model is an explicit supersymmetric extension of the Standard Model,
 which involves a new strongly-interacting $SU(5)$ gauge theory at TeV-scale energies.
We show that the Pentagon can be embedded into an $SU(5) \times
SU(5) \times SU(5)$ gauge group at the GUT scale.  The
doublet-triplet splitting problem, and proton decay compatible with
experimental bounds, can be successfully addressed in this context.
The simplest approach fails to provide masses for the lighter two
generations of quarks and leptons; however, this problem can
 be solved by the addition of a pair of antisymmetric tensor fields and an axion.}

 \received{} \accepted{}
 \preprint{hep-th{07}\\\\scipp-07/13 \\}
 
\begin{document}

 \section{\bf Introduction}
 The Pentagon model\cite{pentagon}\cite{remodel} is the simplest model
 of TeV scale physics, which is compatible with the hypothesis of
 Cosmological SUSY Breaking (CSB)\cite{tbfolly} and with
 phenomenology. The original model\cite{pentagon} relied on a
 complicated singlet sector, which {\it might} have supported a
 meta-stable\footnote{Meta-stability is used in the sense of
 non-gravitational effective field theory.  The arguments of
 \cite{regulate} and \cite{remodel} suggest that the de Sitter
 solution corresponding to this state is as stable as the
 equilibrium configuration of a finite system with a large number of
 degrees of freedom can ever be.  ``Instabilities" occur on the
 recurrence time scale and represent transient fluctuations into low
 entropy states.} SUSY violating state. That model contained a
 light, axion with relatively low decay constant.   It could be made
 barely compatible with experiment, but only by re-introducing the
 strong CP problem.

 The remodeled version of the Pentagon model\cite{remodel} relied
 instead on the arguments of Intriligator Seiberg and
 Shih\cite{iss}(ISS), that SUSY QCD with $N_F = N_C$ and a mass term had a
 meta-stable SUSY violating state\footnote{These arguments have been
 criticized in \cite{yael}.  If we give separate mass to one pentaquark and add
 the coupling to singlets used in the Pentagon model, then the meta-stable
 state exists in one portion of the two parameter phase diagram.
 The question of where the phase boundary is, and whether it extends
 into the region of phenomenological relevance for the Pentagon
 model, cannot be answered by the perturbative methods of
 \cite{yael}.}.   There was no light axion and the singlet sector
 consisted of a single chiral superfield $S$. The model contains a
 scalar pseudo-Goldstone boson, the {\it penton}, stemming from the
 spontaneous breakdown of {\it pentabaryon number}, which is a
 characteristic of the ISS state.   This particle probably evades
 all experimental bounds, but might be discovered in a re-analysis
 of (or further experiments on) flavor changing charged current
 hadron decays.   If the scale at which the accidental pentabaryon
 number symmetry is explicitly broken is between $10^8$ and $10^{10}$
 GeV, the penton field might be responsible for both baryogenesis
 and dark matter\cite{pentabaryogen}. We remind our readers that,
 like most low energy SUSY breaking models, the Pentagon does not
 have a SUSY neutralino dark matter candidate.

 In the present
 paper we will {\it not} assume that the scale of pentabaryon
 number symmetry breaking is in this range. If the symmetry breaking
 scale for pentabaryon number takes the more natural value of the unification
 scale, then the penton {\it might} be the origin of baryogenesis,
 but will make a negligible contribution to the dark matter density.
 The more ambitious program of \cite{pentabaryogen} would require us
 to explain the appearance of the intermediate scale, and to make
 sure that the physics at this scale does not lead to proton decay.
 We will not attempt to construct such a model in this paper.
Indeed, the remodeled Pentagon has a strong CP problem, which we
propose to solve with a QCD axion with large decay constant, $f_a$.
The current (cosmological history independent) upper bound on $f_a$
is of order $10^{14} - 10^{15}$ GeV\cite{pjfstbdg}, and it can
easily be used as a dark matter candidate. We will find that most of
our unification scale models require us to introduce the axion for a
rather different task: the cancellation of discrete anomalies.  Thus,
the scenario suggested by the present paper is that axions are the
dark matter, while the penton might play a role in the generation of
baryon asymmetry. We will call the superfield that contains the
axion $X$.
Most other phenomenological problems of generic SUSY models
 are resolved by the general structure of the
 Pentagon. However, the question of whether it predicts a consistent
 pattern for the electro-weak breaking scale and the super-partner
 spectrum depends on strong coupling physics and does not have a
 definitive answer at this time. The model contains new degrees of
 freedom at the TeV scale so it is not obvious that it has a {\it
 little hierarchy problem}\cite{dst}.

 At the one loop level, the Pentagon model is compatible with
 coupling unification, with a GUT scale coupling that is barely
 perturbative. Dimension $6$ proton decay is probably within reach
 of planned experiments. The purpose of the present paper is to see
 whether the Pentagon model can indeed be embedded in a unified
 model. It is the authors' opinion that the most plausible
 explanation of the discrepancy between the unification and Planck
 scales is that proposed by Witten\cite{wit} in the context of the
 Ho{\hacek r}ava-Witten\cite{horwit} strongly coupled heterotic
 string. In this sort of scenario, quantum gravitational corrections to
 the four dimensional effective field theory of matter\footnote{This term is used
 to distinguish fields whose origin is on a brane, from bulk fields like the
 four dimensional graviton.} are expected to be scaled by the
 unification scale $M_U \sim 2 \times 10^{16}$ GeV.

 It is therefore not strictly correct to use effective field theory
 to describe gauge unification. In this paper, we do this as a
 temporary stopgap measure. It is highly probable that none of the
 currently understood supersymmetric string solutions corresponds to
 the zero cosmological constant (c.c.) limit of CSB\cite{tbfolly}.
 It is absolutely certain that at most one of them does. To make
 progress without making a commitment to a particular string theory
 model, we resort to effective field theory, but do not neglect
 higher order terms in the superpotential\footnote{We will be searching for
 supersymmetric vacua, so we will not need to say anything about the Kahler potential.}.

 Our strategy will be to find a GUT model whose spectrum below
 the GUT scale consists of precisely the fields of the Pentagon
 model. This requires us to solve the doublet-triplet splitting
 problem, and to find an origin of the $SU(3,2,1)$ singlet field of
 the Pentagon (which cannot be a singlet of the unified group). To
 do this, we employ the strategy of \cite{decon}, realizing the
 standard model as part of the diagonal subgroup of an $SU(5) \times
 SU(5)$ gauge group, broken by fields in the $\fivefivebar $ and $\fivebarfive$.
 We will also require that the theory contains an exact $Z_4$ R
 symmetry, which is preserved by the vacuum state. In the philosophy
 of CSB\cite{tbfolly} this symmetry of the effective field theory is
 explicitly broken by interactions of the gravitino with the cosmological horizon\cite{susyhor}
 and the symmetry breaking terms vanish like a power of the c.c.  The
 leading symmetry breaking term induces spontaneous SUSY breaking
 and gives a gravitino mass of order $\Lambda^{1/4}$. In the
 Pentagon model this is the ISS mass term $m_{ISS} P^i_a
 \tilde{P}^a_i $. In previous discussions it has been assumed that
 all other explicit R breaking was a higher power of the c.c., and
 therefore negligible.

In the next section, we construct what we believe is the simplest
model realizing the goals of the above paragraph. However, in section 4 we find
that we cannot reproduce the exact structure of the Pentagon model,
in the sense that we cannot realize the $Z_4$ charge assignments
that were used in \cite{remodel}. This is a consequence of the
intricate requirements imposed by anomaly cancellation, both for the
gauge group and for the $Z_4$ symmetry. As a consequence we find
that we cannot choose the charges to both eliminate dangerous
operators which could lead to proton decay, and allow full rank mass
matrices for quarks and leptons. In \cite{remodel}, the $Z_4$ was
chosen generation blind, but this is impossible in the model we
construct. The problems remain even if we try to impose other
anomaly free discrete symmetries, or use a Green-Schwarz mechanism
involving the field $X$ to cancel some of the discrete anomalies.

Eventually, we traced the problem back to the fact that our model
had an odd number of chiral fields in each of the $SU(5)$ groups.
The simplest way to solve it would be to add an additional
$(\bar{5},1) \oplus (1,5)$ to the model, but this leaves over too
many massless low energy fields. This can be remedied if we add a
$(\overline{10},1) \oplus (1,10)$. We then obtain a model whose low
energy spectrum and $Z_4$ charge assignments agree precisely with
the Pentagon model. To cancel discrete anomalies we have to resort
to a Green-Schwarz mechanism involving $X$. We discuss this model in
Section 5. In the conclusions, we make some comments about the
implementation of the Froggatt-Nielson mechanism in this model, and
about the possibility of allowing R parity violating couplings that
might be useful for resolving the little hierarchy problem\cite{linda}.
Before concluding this introduction, we want to emphasize for
clarity that, although the Pentagon model was motivated by the
highly speculative idea of CSB, it is just a low energy effective
field theory. The only way in which CSB affects any of the analysis
of the Pentagon model is through an {\it a priori} constraint on the
size of the mass parameter $m_{ISS}$.  For readers who prefer to
ignore CSB, one can imagine that this parameter is determined by
{\it retro-fitting}\cite{dfs}.  That is one assumes that it arises
from a non-renormalizable coupling to {\it e.g.} the squared field
strength of a pure supersymmetric gauge theory with scale
$\Lambda_H$. By playing with $\Lambda_H$ one can obtain a value of
$m_{ISS}$ in a phenomenologically acceptable range.

\section{A minimal model}
\subsection{GUT breaking fields and R-charge assignments}

Following \cite{decon}, we introduce an $SU_1 (5) \times SU_2 (5)$
gauge group at the GUT scale, in addition to the $SU_P (5)$ of the
Pentagon. The Standard Model matter fields (three $\bar{5}$'s and
three $10$'s) can each reside in either of these $SU(5)$'s
(transforming as singlets under the other) subject to the constraint
that there are no anomalies in the gauge symmetry. The SM Higgs
fields come from the doublet components of $H_u$ which transforms as
a $(5,1)$, and $H_d$ which transforms as a $(1,\bar{5})$.  We add
two bifundamental fields $\Phi_1$ and $\Phi_2$ that transform as
$(5,\bar{5})$ under the gauge group, and 2 bifundamental fields
$\tilde{\Phi}_1$ and $\tilde{\Phi}_2$ that transform as
$(\bar{5},5)$'s.  These fields will be responsible for breaking the
GUT-scale gauge group.  We adopt the method for solving the
doublet-triplet splitting problem discussed in \cite{decon}. Assume
a SUSic minimum at the following VEVs for the $\Phi$ fields:

$$ \langle\Phi_1 \rangle = \pmatrix {
v_1 & 0  &  0  &  0 & 0 \cr 0   & v_1 &  0  &  0 & 0 \cr 0   & 0   &
v_1  &  0 & 0 \cr 0   & 0  &  0  &  0 & 0 \cr 0   & 0  &  0  &  0 &
0 } \hskip 2cm \langle{\tilde{\Phi}}_1 \rangle = \pmatrix {
\tilde{v_1} & 0  &  0  &  0 & 0 \cr 0   & \tilde{v_1} &  0  &  0 & 0
\cr 0   & 0   &  \tilde{v_1}  &  0 & 0 \cr 0   & 0  &  0  &  0 & 0
\cr 0   & 0  &  0  &  0 & 0 }$$

$$ \langle \Phi_2 \rangle = \pmatrix {
0   & 0  &  0  &  0 & 0 \cr 0   & 0  &  0  &  0 & 0 \cr 0   & 0  &
0  &  0 & 0 \cr 0   & 0  &  0  &  v_2 & 0 \cr 0   & 0  &  0  &  0 &
v_2 } \hskip 2cm \langle\tilde{\Phi}_2 \rangle = \pmatrix { 0   & 0
&  0  &  0 & 0 \cr 0   & 0  &  0  &  0 & 0 \cr 0   & 0  &  0  &  0 &
0 \cr 0   & 0  &  0  &  \tilde{v_2} & 0 \cr 0   & 0  &  0  &  0 &
\tilde{v_2} \cr }$$

If there is a discrete symmetry, which allows a coupling of the
Higgs fields to $\Phi_1$ or $\tilde{\Phi}_1$ but not $\Phi_2$ or
$\tilde{\Phi}_2$, the triplet Higgs will attain a GUT-scale mass
after symmetry breaking while the doublet Higgs will remain
massless.  The $\Phi_2$ fields must get VEVs in the doublet sector
in order to fully break the $SU(5)\times SU(5)$ symmetry down to
$SU(3)\times SU(2) \times U(1)$.  If only the $\Phi_1$ fields got
non-zero VEVs, there would still be an unbroken $SU(2)\times SU(2)$
subgroup.

We also need pentaquark fields, and a field that can couple to the
pentaquarks in such a way that a massless $SU(3,2,1)$ singlet
remains below the GUT scale, which can play the role of the $S$
field in the Pentagon model.  For this
we will introduce a high energy $S$ field which transforms as a $5
\times \bar{5}$ in $SU_{1,2}(5)$, and a $T$ field which transforms
as a 24 in $SU_{1,2}(5)$. (Henceforth, we will refer to the low
energy field of the Pentagon Model, which survives symmetry breaking,
as $s$.)  The plethora of GUT scale fields is needed
in order to ensure that the post GUT spectrum be precisely that of the
Pentagon model. The pentaquark fields $P$ and $\tilde{P}$ must both
be in the same $SU(5)$ group to avoid anomalies, and this must be
the same group in which $S$ transforms in order for there to be a
coupling $SP\tilde{P}$.  They are (anti-)fundamentals in $SU_{1,2}(5)$
and in the Pentagon gauge group, but singlets in $SU_{2,1}(5)$. The
choice of which $SU(5)$ group the T field transforms under
defines two classes of models, which we will later distinguish with
a two valued parameter $p = 0,1$.  Note that the distinction between
the two $SU(5)$ groups is that the high energy avatar of $H_u$
transforms under $SU_1 (5)$.

Finally, we have to impose a $Z_4$ R symmetry, to match that of the
low energy Pentagon model. It is tempting to imagine that this R
symmetry also plays the role of forbidding the unwanted couplings
between the Higgs fields and the $\Phi$ fields. The low energy $R$
symmetry, $Z_4^{\prime}$ may be a combination of the high energy
$Z_4$ with elements of the spontaneously broken GUT group.

Given the VEVs above for the $\Phi$ fields and the following two
requirements:
\begin{list}{}{}
\item{1.} The low energy theory contains a leftover $Z_4'$ R-symmetry after the high energy GUT group is spontaneously broken.
\item{2.} The $SU(2)$ block of components of $\Phi_1$ at low energies have R-charge 2
under the $Z_4'$\footnote{We make this assignment to ensure that the
diagonal (singlet) part of it can mix with the other massless
singlet fields to form the $s$ field of the Pentagon model. $s$ must
have R-charge 2 and needs to contain a piece of $\Phi_1$ in order to
have a coupling to the Higgs.}.
\end{list}

\noindent there is a unique assignment of the R-charges for the
$\Phi$ fields.  There can be a low-energy $Z_4^{\prime}$ preserved
only if there is some combination of $SU(5)$ transformations
combined with the high-energy $Z_4$ transformation that preserves
the VEVs of the $\Phi$ fields (in addition, the VEVs of $S$ and $T$
must be preserved, but we will not impose that just yet).  This can
be accomplished by a simultaneous anti-diagonal $U(1)$ (hypercharge)
rotation in each of the $SU(5)$ groups\footnote{A diagonal rotation
is not useful at present since it is part of the unbroken symmetry
and does not affect the $\Phi$ VEVs.  Combining the specific
antidiagonal transformation we find in this section with different
diagonal transformations yields a class of gauge-equivalent low
energy R-symmetries.}.

$$ \langle \Phi_1 \rangle \rightarrow \exp({2\pi i\over4} q_1) \exp(2i\alpha) \langle\Phi_1\rangle $$
$$ \langle \tilde{\Phi}_1 \rangle \rightarrow \exp({2\pi i\over4} \tilde{q}_1) \exp(-2i\alpha) \langle \tilde{\Phi}_1\rangle $$
$$ \langle \Phi_2 \rangle \rightarrow \exp({2\pi i\over4} q_2) \exp(-3i\alpha) \langle \Phi_2 \rangle$$
$$ \langle \tilde{\Phi}_2 \rangle \rightarrow \exp({2\pi i\over4} \tilde{q}_2) \exp(3i\alpha) \langle \tilde{\Phi}_2\rangle $$

\noindent where $\alpha$ is the angle of an anti-diagonal $U(1)$ rotation in
each of the $SU(5)$'s ($e^{i\alpha}$ in the $SU(3)$ subgroup of
$SU_1(5)$, and $e^{-i\alpha}$ in the $SU(3)$ subgroup of
$SU_2(5)$, with $e^{-{3\over 2}i\alpha}$ and $e^{{3\over
2}i\alpha}$ for the corresponding $SU(2)$ subgroups).  The
constraint that the $\Phi$ VEVs be preserved can then be written:

$$ {q_1\over4} + 2\alpha' = n$$
$$ {\tilde{q}_1\over4} - 2\alpha' = m$$
$$ {q_2\over4} - 3\alpha' = l$$
$$ {\tilde{q}_2\over4} + 3\alpha' = r$$

\noindent where $\alpha' \equiv {\alpha\over 2\pi}$, and $n,m,l,r$ are
integers.  Combining the first two equations requires that
$\tilde{q}_1 = -q_1$, and the last two equations imply that
$\tilde{q}_2 = -q_2$.  In other words, the tilded fields must have
 R-charge opposite that of the corresponding untilded fields. Since the $SU(2)$ block of $\Phi_1$
 transforms as $\exp({2\pi i\over 4} q_1) \exp(-3i\alpha)$, the constraint coming
from condition 2 above is:

$$ {q_1\over 4} - 3\alpha' = 1/2 + j $$

\noindent where $j$ is another independent integer.  Combining this with the
first and third equations from constraint 2 implies that $q_1 = 0$
mod 4, $q_2 = 2$ mod 4, and $\alpha' = {1\over 2}$ mod 1 (or $\alpha
= \pi$).  Therefore, the only assignment of high-energy R-charges
that is compatible with the two requirements is $R(\Phi_1) = 0$,
$R(\tilde{\Phi}_1) = 0$, $R(\Phi_2) = 2$, $R(\tilde{\Phi}_2) = 2$.

The $U(1)$ transformation which yields $Z_4'$ when combined with
$Z_4$ is given by:

$$ G = \pmatrix {
-1 & 0  &  0  &  0 & 0 \cr 0   & -1 &  0  &  0 & 0 \cr 0   & 0   &
-1  &  0 & 0 \cr 0   & 0  &  0  &  i & 0 \cr 0   & 0  &  0  &  0 & i
\cr }$$

\noindent where the transformation we need is ``anti-diagonal" in
the sense that we combine $G$ acting on $SU_1(5)$ with $G^{\dagger}$
acting on $SU_2(5)$.  The fields then transform like $\Phi_i
\rightarrow G\Phi_i G$, $\tilde{\Phi}_i \rightarrow
G^{\dagger}\tilde{\Phi}_i G^{\dagger}$, $S \rightarrow G S
G^{\dagger}$, and $T \rightarrow G T G^{\dagger}$. The components in
the $SU(3)$ block of all of the $\Phi$ fields are even under the
$U(1)$ rotation, so they have the same $Z_4'$ charges as their $Z_4$
charges.  However, the components in the $SU(2)$ block of all of the
$\Phi$ fields are odd under the $U(1)$ rotation, so their $Z_4'$
charges are opposite to their $Z_4$ charges.  The components in the
off-diagonal blocks of the $\Phi_i$ fields transform as
$\Phi_i^{(3,2)} \rightarrow -i\Phi_i^{(3,2)}$ (and the same for
$\Phi_i^{(\bar{3},2)}$, where the superscript is referring to their
transformation properties under the appropriate $SU(3) \times SU(2)$
subgroup), so their $Z_4'$ R-charges are their $Z_4$ charges minus
1.  The $\tilde{\Phi}_i^{(3,2)}$ and $\tilde{\Phi}_i^{(\bar{3},2)}$
components transform oppositely to this, so their $Z_4'$ charges are
their $Z_4$ charge plus 1.

For the $S$ and $T$ fields, the on-diagonal blocks are both
invariant under the hypercharge transformation, so their R-charges
are unchanged after the symmetry is broken.  Assuming the $S$ and
$T$ fields are in $SU_1(5)$, the R-charges of the $(3,2)$
(lower-left) blocks of the $S$ and $T$ fields are decreased by 1,
whereas the R-charges of the $(\bar{3},2)$ (upper-right) blocks of
the $S$ and $T$ fields are increased by 1.  This assignment is
reversed if one or both of the fields is in $SU_2(5)$. Note that the
$(3,2)$ numbers here refer to charges under the $SU(3) \times SU(2)$
subgroup of $SU_1(5)$, whereas in the previous paragraph
(for the $\Phi$'s) they are referring to a product of the $SU(3)$
subgroup from one $SU(5)$ and the $SU(2)$ subgroup from the other
$SU(5)$.

In order to preserve the R-symmetry, the VEVs of the $S$ and $T$
fields must also be invariant under the $Z_4'$ transformation. The
on-diagonal blocks of $S$ and $T$ are already invariant, but the
off-diagonal blocks are not.  Therefore, the off-diagonal blocks
must have zero VEVs.  In order to have a coupling between $S$ and
the pentaquark fields, we need the entire VEV of $S$ to be zero to
avoid giving the pentaquarks a GUT scale mass. Therefore, the only
allowed non-zero VEVs for the $S$ and $T$ fields are in the diagonal
blocks of $T$.  After constructing the superpotential, we will need
to verify that it has a minimum where this is the case.

\begin{table}[!h!t!b]
\begin{center}
\begin{tabular}{|c|c|c|c|c|c|c|}
\hline
 & $SU(5) \times SU(5)$ & $Z_4$ & $(8,1)$ $Z_4'$ & $(1,3)$ $Z_4'$ & $(3,2)$ $Z_4'$ & $(\bar{3},2)$ $Z_4'$ \\
\hline
$\Phi_1$ & $(5,\bar{5})$ & 0 & 0 & 2 & 3 & 3 \\
\hline
$\tilde{\Phi}_1$ & $(\bar{5},5)$ & 0 & 0 & 2 & 1 & 1 \\
\hline
$\Phi_2$ & $(5,\bar{5})$ & 2 & 2 & 0 & 1 & 1\\
\hline
$\tilde{\Phi}_2$ & $(\bar{5},5)$ & 2 & 2 & 0 & 3 & 3 \\
\hline
$S$ & $(5\times \bar{5},1)$ & 2 & 2 & 2 & $1^*$ & $3^*$ \\
\hline
$T$ & $(24,1)$ & 0 & 0 & 0 & $3^*$ & $1^*$ \\
\hline
\end{tabular}
\end{center}
\caption{High and low energy R-charges of GUT-breaking fields}
\label{GUTfieldRchargeTable}
\end{table}

Table~\ref{GUTfieldRchargeTable} summarizes the $\Phi$, $S$, and $T$ fields, their $Z_4$ charges, and the low-energy $Z_4'$ charges of the different components.  If the $S$ is a $(1,5\times \bar{5})$ or the $T$ is a $(1,24)$ then
the 1's and 3's (marked by $*$'s) in the corresponding row are
reversed.  There are also $(1,1)$ singlet components in all of the
fields, which are not listed in the table.  There are two in each of
the $\Phi$ fields:  one is the trace of the $SU(3)$ block and has
the same R-charge as the $(8,1)$ components, and the other is the
trace of the $SU(2)$ block and has the same R-charge as the $(3,1)$
components.  Similarly, there are also two $(1,1)$ components in the
$S$ which have the same R-charge as the rest of the diagonal blocks.
And finally, there is one $(1,1)$ component in the $T$:  the
component proportional to the generator of $SU(5)$ responsible for
$U(1)$ hypercharge transformations in the usual embedding of $SU(3)
\times SU(2) \times U(1)$.  In all of these cases, the $Z_4'$ charge
of the singlets is just the same as the corresponding $(8,1)$ or
$(3,1)$ components in the table above.

The low-energy R-charges of the matter fields of the Standard Model,
Higgs fields, and pentaquarks are also different from their
high-energy R-charges.  Both of the Higgs doublets end up with their
low-energy R-charge increased by 1.  Since there is a high-energy
coupling ${\tilde \Phi}_1 H_u H_d$ (this is the term which gives the
triplet Higgses a GUT-scale mass when the VEV of $\Phi_1$ is plugged
in), the sum of the Higgs $Z_4$ charges must be 2.  At low energies
(in the Pentagon model) the R-charge of both Higgs fields must be
opposite in order to couple to the low-energy $S$ field.  This works
out automatically, since the symmetry breaking changes their total
R-charge by 2.

The pentaquarks are more complicated, since none of their components
get masses after the breaking.  Different components of the
pentaquarks have different R-charges at low energy. However, this
is equivalent to the R-symmetry described in the original Pentagon
model in which there is no distinction between the R-charge of
doublet and triplet components. The argument goes as follows: The
R-charges of the pentaquark fields start out summing to zero, and
when they get broken down they must end up summing to zero. This is
automatically the case for both the doublet and triplet components,
since the change in R-charge upon breaking is opposite for a $(5,1)$
as it is for a $(\bar{5},1)$. Because $P$ and $\tilde{P}$ are the
only fields charged under the {\it Pentagon} $SU(5)$, they only show
up in pentabaryon and pentameson combinations in the low-energy
superpotential. $P\tilde{P}$ still has R-charge 0, and $\det[PPPPP]$
(and similarly $\det[{\tilde P}{\tilde P}{\tilde P}{\tilde P}{\tilde
P}]$) is a product of 2 doublet P's and 3 triplet P's that adds up
to R-charge 0 at low energies as well. Consequently, all of the
terms involving $P$ and $\tilde{P}$ in the original Pentagon model
are present and no additional terms are present.  In other words,
the $Z_4'$ R-symmetry described here differs from the one in the
original Pentagon model only by a transformation which is already a
symmetry of the superpotential--in particular, a $U(1)$ hypercharge
transformation. Instead of using the anti-diagonal transformation
$G$ to relate the high and low energy R-symmetries, we could have
chosen to perform the transformation twice in one $SU(5)$ (leaving
the pentaquark fields in the other $SU(5)$ invariant). This
transformation differs from G only by an element of the unbroken
(diagonal) $U(1)$.

The Standard Model matter fields break down into $\bar{5} =
(\bar{3},1) \oplus (1,2)$ ($\bar{D}$ and $L$), and $10 = (3,2)
\oplus (\bar{3},1) \oplus (1,1)$ ($Q$, $\bar{U}$, and $\bar{E})$. In
the original formulation of the Pentagon model, it was assumed that
all 3 generations of $\bar{D}$ had R-charge $2+3R(L)-R(H_d)$.
However, this is inconsistent with our present description of $L$
and $\bar{D}$ as coming from a $\bar{5}$ field with a single
R-charge at high energies.  After GUT breaking, the R-charge of $L$
changes by 1, whereas the R-charge of $\bar{D}$ changes by 2.
Therefore, for a particular generation one of them must have an odd
R-charge and the other an even R-charge.  Furthermore, in our present setup
anomaly cancellation requires us to place different matter fields in
different $SU(5)$ groups, which makes it necessary to drop the assumption
made in the original Pentagon model that all three generations have the
same R-charge at low energies.  This also suggests that the high-energy
R-charges may be different for each generation. We will consider all
possibilities for both the high energy R-charges and the high energy
$SU_1(5) \times SU_2 (5)$ quantum numbers of the Standard Model
matter fields.  In order to stay within experimental bounds for
lepton and baryon violation, it will also be necessary to add at
least one additional discrete symmetry to the high energy model.
Table~\ref{SMHiLowRCharge} summarizes the way the R-charges of the
Standard Model fields change after symmetry breaking.  The value of
each R-charge is abbreviated here by the corresponding field (a
convention we will adhere to from now on in this paper). Depending
on the generation, each field of a given type may transform either
in $SU_1(5)$ or in $SU_2(5)$ (each choice
corresponding to a column in the table), except for the Higgs fields
whose transformation properties are fixed by our choice $SU_1 (5)$
as the group under which the $H_u$ transforms.
\begin{table}[!h!t!b]
\begin{center}
\begin{tabular}{|c|c|c|}
\hline

& $SU_1(5)$ & $SU_2(5)$ \\
\hline
L  & ${\cal D} - 1$ & ${\cal D} + 1$ \\
$\bar{D}$ & ${\cal D} + 2$ & ${\cal D} + 2$ \\
Q & ${\cal U} - 1$ & ${\cal U} + 1$ \\
$\bar{U}$ & ${\cal U}$ & ${\cal U}$  \\
$\bar{E}$ & ${\cal U} + 2$ & ${\cal U}+2$ \\
$h_u$ & $H_u + 1$ & \\
$h_d$ &           & $H_d + 1$ \\
\hline
\end{tabular}
\end{center}
\caption{High and low energy R-charges of Standard Model fields}
\label{SMHiLowRCharge}
\end{table}

Knowing the R-charges of all the low energy field components allows
us to explain our choice of high energy fields.  The $\Phi$ fields
exist at the GUT scale to provide the mechanism for breaking to the
$SU(3) \times SU(2) \times U(1)$ of the Standard Model and give
the triplet Higgs a mass, but they are not present in the Pentagon
model and so must gain a mass.  However, the model does have a
singlet ($s$), which couples to both the Pentaquarks and the Higgs
doublets. This field must originate from fields at the GUT scale
with the same couplings but remain massless at low energies.
$\tilde{\Phi}_1$ couples to the Higgs, but cannot couple to the
Pentaquarks because it would give them a mass; this is why it was
necessary to introduce a field with zero VEV ($S$) with such a
coupling.  However, the Pentaquarks both transform in a single $SU(5)$
while the Higgs must couple to fields that transform under both
groups. Therefore, the Pentagon singlet $s$ must be a massless
linear combination of the singlet components of the high energy
fields, and must include the singlets of $S$ as well as the $SU(2)$
block singlet of $\Phi_1$ (at low energies the $s$ need only couple
to the Higgs doublet). All other components of the high energy
fields must acquire a mass to prohibit them from appearing in the
low energy model.

The inclusion of $S \in (5 \times \bar{5},1)$ and $T \in (24,1)$ in the
high energy model is necessary to provide mass to all unwanted fields.
In total, $S,T,\Phi_1,\tilde{\Phi}_1,\Phi_2,\tilde{\Phi}_2$ comprise
149 components; 148 must acquire mass while one remains massless.
Finding exactly one zero eigenvalue by diagonalizing a 149 row
$\times$ 149 column mass matrix is not a trivial task, but this
naive approach is greatly simplified by two facts: every mass term
in the low energy effective lagrangian must have R-charge 2 and must
exhibit $SU(3) \times SU(2) \times U(1)$ invariance.  The group
structure of the theory allow us to split the mass matrix into group
irreducible blocks; in particular, the eleven singlets would form an
11$\times$11 block, the $SU(2)$ components form an 18$\times$18 block,
the $SU(3)$ components a 48$\times$48 block, and the (3,2) and
($\bar{3},2$) components would combine into two identical
36$\times$36 blocks.  Each block must have all non-zero eigenvalues
except for one corresponding to an eigenvector from the singlet
block with components in the direction of $S$ and the $SU(2)$ block
singlet of $\tilde{\Phi}_1$.

This is indeed the case for our model, ensured by the R-charges of
the fields.  In the $SU(2)$ and $SU(3)$ sectors there are an equal
number of R-charge 2 and R-charge 0 components, allowing every one
of these components to 'pair up' into quadratic mass terms. Thus
every (3,1) and (1,2) component of the high energy fields will
gain a GUT scale mass. The argument is similar for the (3,2) and
($\bar{3}$,2) components: there is exactly one R-charge 3 (3,2)
component for every R-charge 3 ($\bar{3}$,2) (the same is true
for the R-charge 1 components), again allowing each of these fields
to gain mass.  This is not the case for the singlet block--there are
an odd number of fields, six with R-charge 2 and five with R-charge
0, so there will inevitably be a single R-charge 2 field that cannot
pair with any other field and will therefore remain massless.
Furthermore, in the general case this field will be some linear
combination of all of the R-charge 2 singlets, including the
singlets of the $S$ and the $SU(2)$ block singlet of $\tilde{\Phi}_1$.

The choice of GUT scale field content in our model is not unique.
The requirement of obtaining one massless singlet in the low energy
limit, with all the rest of the fields massive, enforces two
conditions.  First, there must be an odd number of singlet
components, $(N + 1)/2$ of which are R-charge 2 and the rest
R-charge zero. This field will in general be a linear combination of
all the R-charge 2 components, and so, assuming the additional
requirement of the high energy couplings $S P \tilde P$ and
$\tilde{\Phi}_1 H_u H_d$ as discussed, this singlet will include
components in the directions of both $S$ and the $SU(2)$ block
singlet of $\tilde{\Phi}_1$. The second condition is that the total
number of fields is even, half with R-charge 2 and half with
R-charge 0, so that all components other than the singlets can pair
with another field and so gain mass.  Our particular choice of
fields is the simplest we have found, which solves the
doublet-triplet splitting problem while having low energy field
content identical to the Pentagon model.

\section{The Superpotential}

Thus far we have constructed a GUT scale $SU(5) \times SU(5)$
$\times$ $SU(5)$ model that spontaneously breaks to the Pentagon $SU(5)$
$\times$ Standard Model $SU(3) \times SU(2) \times U(1)$. It
remains to construct the superpotential of this model. In
particular, we must show that our VEVs lie at a minimum of the
potential and that the mass spectrum of the low energy effective
theory does in fact leave the field content of the Pentagon model.
In this section we will focus on the additional ingredients that our
high energy model has contributed to the Pentagon, assuming that all
other fields have zero GUT scale VEVs and so do not contribute to
the discussion.  A more detailed consideration of the standard
matter fields will be the subject of the next section.

All terms in the superpotential must obey two rules: they must be
invariant under both of the $SU(5)$ symmetries (here and for the
majority of the discussion we will ignore the Pentagon $SU(5)$ as all
fields but the pentaquarks transform trivially under it), and the
total R-charge of each term must sum to 2 .  Because all of the
fields under consideration have either R-charge 2 ($S,\Phi_2,
\tilde{\Phi}_2$) or R-charge 0 ($T, \Phi_1, \tilde{\Phi}_1$), there
must be an odd number of R-charge 2 fields in every term.  The $SU(5)$
invariance has a number of consequences.  We will take the group
structure of the fields under $SU_1(5) \times SU_2(5)$ to be $S \in (5
\times \bar{5},1), T \in (24,1), \Phi_i \in (5,\bar{5}),$ and
$\tilde{\Phi}_i \in (\bar{5},5)$.  The discussion is analogous for the
case that the $S$ and/or $T$ transform in $SU_2(5)$.   Then,
since only the bi-fundamentals transform under $SU_2(5)$,
gauge invariance implies that there must be an even number of them
in every term, with a $\Phi_i$ always paired with a
$\tilde{\Phi}_j$.  These combined pairs of $\Phi \tilde{\Phi}$
behave as a single $5 \times \bar{5}$ field under $SU_1(5)$,
and can either be traced over or combined with the other fields in
terms that are invariant under $SU_1(5)$.

First look at terms that only involve traces over the $\Phi$s, we
will call this piece of the superpotential $W_{\Phi}$,

\begin{eqnarray*}
W_{\Phi} & = & M (\Phi_1 \tilde{\Phi}_2) + M (\Phi_2 \tilde{\Phi}_1) \\
& & + {1 \over M}[(\Phi_1 \tilde{\Phi}_1 \Phi_1 \tilde{\Phi}_2) +
(\Phi_1 \tilde{\Phi}_1 \Phi_2 \tilde{\Phi}_1) + (\Phi_1
\tilde{\Phi}_2 \Phi_2 \tilde{\Phi}_2) + (\Phi_2 \tilde{\Phi}_1
\Phi_2 \tilde{\Phi}_2)
\\
& & + (\Phi_1 \tilde{\Phi}_1) ( \Phi_1 \tilde{\Phi}_2) + (\Phi_1
\tilde{\Phi}_1) ( \Phi_2 \tilde{\Phi}_1) + (\Phi_1 \tilde{\Phi}_2) (
\Phi_2 \tilde{\Phi}_2) + (\Phi_2 \tilde{\Phi}_1)( \Phi_2
\tilde{\Phi}_2)]
\\ & & + \hbox{\rm higher order.}
\end{eqnarray*}

We have suppressed the $SU(5)$ index structure. Terms in parenthesis
imply a trace over those fields.  So for example, the term $(\Phi_1
\tilde{\Phi}_1 \Phi_1 \tilde{\Phi}_2)$ would be written explicitly
as $(\Phi_1)^i_A (\tilde{\Phi}_1)^A_j (\Phi_1)^j_B
(\tilde{\Phi}_2)^B_i$, whereas the term $(\Phi_1 \tilde{\Phi}_1) (
\Phi_1 \tilde{\Phi}_2)$ would be $(\Phi_1)^i_A (\tilde{\Phi}_1)^A_i
( \Phi_1)^j_B (\tilde{\Phi}_2)^B_j$.  An upper index refers to the
$5$ and a lower index refers to the $\bar{5}$ representation, while
the lower case indices refer to $SU_1(5)$ and the upper case
indices to $SU_2(5)$.  We have also omitted coefficients for
the terms, which in general should be arbitrary.

The mass scale appearing in these equations is of order the GUT
scale, which according to our hypothesis, is the scale at which we
expect quantum gravitational corrections to appear. Our strategy
will be to work with polynomials of minimal order, to demonstrate
that we can achieve the pattern of VEVs we used in the previous
section.  The low order terms do not have accidental symmetries, so
we expect that higher order corrections will make order one changes
to the VEVs and masses, without disturbing the qualitative nature of
the system or the low energy Lagrangian.

There will be other terms in $W$ involving the determinants of the
$\Phi$ fields.  We will separate these onto their own:

\begin{eqnarray*}
W_{det} & = & M^{-2} (\det [ \Phi_1 \Phi_1 \Phi_1 \Phi_1 \Phi_2] +
\det [ \Phi_1 \Phi_1 \Phi_2 \Phi_2 \Phi_2] +  \det [\Phi_2
\Phi_2 \Phi_2 \Phi_2 \Phi_2]
\\
& & + \det [ \tilde{\Phi}_1 \tilde{\Phi}_1 \tilde{\Phi}_1
\tilde{\Phi}_1 \tilde{\Phi}_2] + \det [ \tilde{\Phi}_1
\tilde{\Phi}_1 \tilde{\Phi}_2 \tilde{\Phi}_2 \tilde{\Phi}_2] +
\det [ \tilde{\Phi}_2 \tilde{\Phi}_2 \tilde{\Phi}_2 \tilde{\Phi}_2
\tilde{\Phi}_2)]
\\ & &
+ \hbox{\rm higher order.}
\end{eqnarray*}

Next consider the terms involving the $S$ and $T$ fields,

\begin{eqnarray*}
W_{S} & = & (S \Phi_1 \tilde{\Phi}_1) + (S \Phi_2 \tilde{\Phi}_2)
\\ & &
 + {1 \over M} [(S S \Phi_1 \tilde{\Phi}_2) + (S S \Phi_2
\tilde{\Phi}_1) + (S S) (\Phi_1 \tilde{\Phi}_2) + (S S) (\Phi_2
\tilde{\Phi}_1)]
\\ & &
 + {1 \over M^2} [ (S \Phi_1 \tilde{\Phi}_1
\Phi_1 \tilde{\Phi}_1) + (S \Phi_2 \tilde{\Phi}_2 \Phi_2
\tilde{\Phi}_2) + (S \Phi_1 \tilde{\Phi}_1 \Phi_2 \tilde{\Phi}_2) +
(S \Phi_1 \tilde{\Phi}_2 \Phi_2 \tilde{\Phi}_1)
\\ & &
+ (S \Phi_2 \tilde{\Phi}_2 \Phi_1 \tilde{\Phi}_1) + (S \Phi_1
\tilde{\Phi}_2 \Phi_1 \tilde{\Phi}_2) + (S \Phi_2 \tilde{\Phi}_1
\Phi_2 \tilde{\Phi}_1) + (S \Phi_2 \tilde{\Phi}_1 \Phi_1
\tilde{\Phi}_2)
\\ & &
 + (S \Phi_1 \tilde{\Phi}_1) ( \Phi_1 \tilde{\Phi}_1) + (S \Phi_2
\tilde{\Phi}_2) ( \Phi_2 \tilde{\Phi}_2) + (S \Phi_1 \tilde{\Phi}_1)
( \Phi_2 \tilde{\Phi}_2) + (S \Phi_1 \tilde{\Phi}_2) ( \Phi_2
\tilde{\Phi}_1)
\\ & &
+ (S \Phi_2 \tilde{\Phi}_2) ( \Phi_1 \tilde{\Phi}_1) + (S \Phi_1
\tilde{\Phi}_2) ( \Phi_1 \tilde{\Phi}_2) + (S \Phi_2 \tilde{\Phi}_1)
( \Phi_2 \tilde{\Phi}_1) + (S \Phi_2 \tilde{\Phi}_1) ( \Phi_1
\tilde{\Phi}_2)]
\\ & &
+ \hbox{\rm higher order},
\end{eqnarray*}

\begin{eqnarray*}
W_{T} & = & (T \Phi_1 \tilde{\Phi}_2) + (T \Phi_2 \tilde{\Phi}_1)
\\ & &
 + {1 \over M} [(T T \Phi_1 \tilde{\Phi}_2) + (T T \Phi_2
\tilde{\Phi}_1) + (T T) (\Phi_1 \tilde{\Phi}_2) + (T T) (\Phi_2
\tilde{\Phi}_1)]
\\ & &
 + {1 \over M^2} [ (T \Phi_1 \tilde{\Phi}_2
\Phi_1 \tilde{\Phi}_1) + (T \Phi_1 \tilde{\Phi}_2 \Phi_2
\tilde{\Phi}_2) + (T \Phi_2 \tilde{\Phi}_1 \Phi_1 \tilde{\Phi}_1) +
(T \Phi_2 \tilde{\Phi}_1 \Phi_2 \tilde{\Phi}_2)
\\ & &
 + (T \Phi_1 \tilde{\Phi}_1 \Phi_1 \tilde{\Phi}_2) + (T \Phi_2 \tilde{\Phi}_2
\Phi_1 \tilde{\Phi}_2) + (T \Phi_1 \tilde{\Phi}_1 \Phi_2
\tilde{\Phi}_1) + (T \Phi_2 \tilde{\Phi}_2 \Phi_2 \tilde{\Phi}_1)
\\ & &
 + (T \Phi_1 \tilde{\Phi}_2) ( \Phi_1 \tilde{\Phi}_1) + (T \Phi_1
\tilde{\Phi}_2) ( \Phi_2 \tilde{\Phi}_2) + (T \Phi_2 \tilde{\Phi}_1)
( \Phi_1 \tilde{\Phi}_1) + (T \Phi_2 \tilde{\Phi}_1) ( \Phi_2
\tilde{\Phi}_2)
\\ & &
+ (T \Phi_1 \tilde{\Phi}_1) ( \Phi_1 \tilde{\Phi}_2) + (T \Phi_2
\tilde{\Phi}_2) ( \Phi_1 \tilde{\Phi}_2) + (T \Phi_1 \tilde{\Phi}_1)
( \Phi_2 \tilde{\Phi}_1) + (T \Phi_2 \tilde{\Phi}_2) ( \Phi_2
\tilde{\Phi}_1)
\\ & &
+ (T T T \Phi_1 \tilde{\Phi}_2) + (T T T \Phi_2 \tilde{\Phi}_1) + (T
T T) (\Phi_1 \tilde{\Phi}_2) + ( T T T) (\Phi_2 \tilde{\Phi}_1)
\\ & &
+ (T T) ( T \Phi_1 \tilde{\Phi}_2) + (T T) ( T \Phi_2
\tilde{\Phi}_1)] + \hbox{\rm higher order.}
\end{eqnarray*}

and

\begin{eqnarray*}
W_{ST} & = & M (S T) + (S T T) + {1 \over M} [(S T \Phi_1
\tilde{\Phi}_1) + (S T \Phi_2 \tilde{\Phi}_2)
\\ & &
 + (T S \Phi_1 \tilde{\Phi}_1) + (T S \Phi_2 \tilde{\Phi}_2) + (S T) (\Phi_1
\tilde{\Phi}_1) + (S T) (\Phi_2 \tilde{\Phi}_2)]
\\ & &
 + \hbox{\rm higher order.}
\end{eqnarray*}

The total superpotential, $W$, will include the sum of these pieces
as well as contributions from terms containing the pentaquarks and
matter fields, $$W = W_{\Phi} + W_{det} + W_S + W_T + W_{ST} +
W_{Pentagon}.$$  Although we will not discuss in detail the content
of $W_{Pentagon}$, we should point out a few of the key terms
mentioned earlier.  Most importantly, the couplings $S P \tilde{P}$
and $H_u \tilde{\phi}_1 H_d$ lead to $W_S$ of the original Pentagon
model. There will also be Yukawa terms including the fields $H_u
{\cal U}_i {\cal U}_j$ and $H_d {\cal U}_i {\cal D}_j$ where the
$\cal{U},\cal{D}$ transform as $10,\bar{5}$ respectively under one
of the $SU(5)$s; these will be discussed further in the next section.

Now that we have constructed the superpotential, we should verify
that the VEVs we chose in the previous section are in fact at a
minimum. We can expect that this will be achieved only by satisfying
a set of six constraints--there are five degrees of freedom in the
VEVs, each of which should be determined by the $F$-equations, and a
sixth constraint will restrict the coefficients of the terms in the
lagrangian in a manner required by the preservation of the R
symmetry.

Let us assume that the vacuum expectation values of the $\Phi$
fields have the form discussed in the previous section, and that the
VEV of $S$ is zero.  The form of the VEV of $T$ is so far
undetermined, but we know from the previous section that it must be
block diagonal.  Furthermore, we will see presently that all off
diagonal components within these blocks will have to be zero as
well, that the $SU(3)$ diagonal components must all be equal to each
other (the same being true for the $SU(2)$ components), and that the
trace must be zero.  Thus we will assume the VEV of $T$ has the form
$$
 \langle T \rangle = \pmatrix {
v_T & 0  &  0  &  0 & 0 \cr 0   & v_T &  0  &  0 & 0 \cr 0   & 0   &
v_T  &  0 & 0 \cr 0   & 0  &  0  &  -3/2 v_T & 0 \cr 0   & 0  &  0
&  0 & -3/2 v_T \cr }. $$

Consider first the $S$ equation, $F_{S} = \partial W / \partial S
=0$. Even before inserting the VEVs, the only surviving terms are
those from $W_S$ and $W_{ST}$. Terms that involve the product of the
VEVs of $\Phi_1$ or $\tilde{\Phi}_1$ times $\Phi_2$ or
$\tilde{\Phi}_2$ in $W_S$, and those with multiple powers of $S$ in
$W_{ST}$, will vanish. What remains is (again, omitting the explicit
index structure; a system of equations for the individual components
is implied):

\begin{eqnarray*}
F_{S} & = & \Phi_1 \tilde{\Phi}_1 + \Phi_2 \tilde{\Phi}_2 + {1 \over
M^2} [ \Phi_1 \tilde{\Phi}_1 \Phi_1 \tilde{\Phi}_1 + \Phi_2
\tilde{\Phi}_2 \Phi_2 \tilde{\Phi}_2
\\ & &
 + \Phi_1 \tilde{\Phi}_1 ( \Phi_1 \tilde{\Phi}_1) +  \Phi_2
\tilde{\Phi}_2 ( \Phi_2 \tilde{\Phi}_2) +  \Phi_1 \tilde{\Phi}_1 (
\Phi_2 \tilde{\Phi}_2) + \Phi_2 \tilde{\Phi}_2 (\Phi_1
\tilde{\Phi}_1) ]
\\ & &
 + M T + T T + {1 \over M} [T \Phi_1
\tilde{\Phi}_1 + T \Phi_2 \tilde{\Phi}_2 +
\\ & &
 \Phi_1
\tilde{\Phi}_1 T + \Phi_2 \tilde{\Phi}_2 T + T (\Phi_1
\tilde{\Phi}_1) + T (\Phi_2 \tilde{\Phi}_2)]
\\ & &  + \hbox{\rm higher order.}
\end{eqnarray*}

\noindent We are really taking derivatives with respect to
individual components of the $S^i_j$, so terms not in parenthesis
should be read as $(\Phi_1)^j_A (\tilde{\Phi}_1)^A_k T^k_i$.  Since
all of the VEVs are diagonal, and preserve $SU(2,3)$, we can
separate $F_S = 0$ into two distinct constraint equations. Until now
we have neglected to include coefficients in front of each of the
terms in the lagrangian but in general they should be arbitrary, so
the resulting constraint equations will have the form

$$
A v_1 \tilde{v}_1 + B (v_1 \tilde{v}_1)^2 + C (v_1 \tilde{v}_1)(v_2
\tilde{v}_2) + D v_T + E v_T^2 + F v_T v_1 \tilde{v}_1 + ... = 0
$$

and

$$
G v_2 \tilde{v}_2 + H (v_2 \tilde{v}_2)^2 + I (v_1 \tilde{v}_1)(v_2
\tilde{v}_2) + J v_T + K v_T^2 + L v_T v_2 \tilde{v}_2 + ... = 0
$$

Notice that had we not chosen the VEV of $T$ to be diagonal, $F_S =
0$ would enforce this to be true, as there are no other terms
present in the off diagonal component equations.  Also note that had
we not chosen the diagonal components of the VEV to be equal, we
would not have been able to split the constraints into two blocks as
we have done above.  Instead, we would have five independent
equations, the equations within each block differing from each other
only by the components replacing $v_T$.  This would force these
components to be equal.

The $F_{T} = 0$ equation is satisfied automatically by our choice of
VEVs.  Parallel to $F_{S}$, the only terms in the $F_{T}$ equation
come from $W_T$ and $W_{ST}$.  The terms from the latter are all
zero due to the zero VEV of $S$, while the terms from the former
must contain an even number of $\Phi$ fields but an odd number
$\Phi_2$ or $\tilde{\Phi}_2$s and so will inevitably involve a
product of $\Phi_1$ or $\tilde{\Phi}_1$ times $\Phi_2$ or
$\tilde{\Phi}_2$.

Each of the $F_{\Phi}$ equations introduces a new constraint.  The
zero VEV of $S$ will eliminate all terms in the $F$-equations from
$W_S$ and $W_{ST}$.  The terms from $W_{det}$ will vanish because
each term will involve a product with at least one zero.   This
leaves $W_{\Phi}$ and $W_{T}$; since these contain similar terms for
each of the $\Phi$s, we will focus in particular on $F_{\Phi_1}$ for
illustration:

\begin{eqnarray*}
F_{\Phi_1} & = & M \tilde{\Phi}_2 + \tilde{\Phi}_2 T + {1 \over
M}[\tilde{\Phi}_2 \Phi_2 \tilde{\Phi}_2 + \tilde{\Phi}_2 (\Phi_1
\tilde{\Phi}_1) + \tilde{\Phi}_2 ( \Phi_2 \tilde{\Phi}_2)]
\\ & &
+ {1 \over M^2}[ \tilde{\Phi}_2 \Phi_2 \tilde{\Phi}_2 T +
\tilde{\Phi}_2 T (\Phi_1 \tilde{\Phi}_1) + \tilde{\Phi}_2 T ( \Phi_2
\tilde{\Phi}_2) + \tilde{\Phi}_2  (T \Phi_1 \tilde{\Phi}_1) +
\tilde{\Phi}_2 (T \Phi_2 \tilde{\Phi}_2)
\\ & & + \hbox{\rm higher order.}
\end{eqnarray*}

Notice that the only surviving terms are proportional to some power
of the VEV of $\tilde{\Phi}_2$ (this is true to all orders).  Thus
the bottom two diagonal components are identical to each other while
the rest of the components are zero, the result of which is a single
equation imposing some new constraint on the $v$s, of the form

$$
0 = A \tilde{v}_2 + B v_2 \tilde{v}_2^2 + C \tilde{v}_2 v_1
\tilde{v}_1 + D v_T \tilde{v}_2 + E v_T v_2 \tilde{v}_2^2 + F v_T
\tilde{v}_2 v_1 \tilde{v}_1 + ...
$$

In addition to the two constraints we have from $F_S =0$, here is a
third equation involving all five degrees of freedom that must be
satisfied in order for our chosen VEVs to lie at a minimum of the
potential.  The equations $F_{\tilde{\Phi}_1} = 0, F_{\Phi_2} =0$
and $F_{\tilde{\Phi}_2} = 0$ each impose an additional new
constraint, all having a similar form to that written above for
$\Phi_1$.  As expected, we end up with six constraints for five
unknowns. This is always the case for a vacuum which preserves both
SUSY and an $R$ symmetry.  Since our VEVs were designed to preserve
an $R$ symmetry, all six constraints are satisfied.

Let us now examine the mass spectrum of the low energy theory.
Spontaneous breaking of the symmetry allows us to re-express the
lagrangian by expanding the fields about the minimum of the
potential, that is $\Phi \rightarrow \langle \Phi \rangle + \phi$,
etc.  The masses of the low energy fields are found by examining the
coefficients of the terms quadratic in the fields, but we are not
concerned with the specific value of the masses, as all masses will
be of the order of the GUT scale and so will be integrated out.

Let us instead simply consider the various components that survive
at low energies after inserting the VEVs.  It is convenient to write
the components of the fields appearing at low energies in terms of
the Gell-Mann basis for the adjoint representation: $S^i_j = S^a
(\lambda^i_j)_a, a = 1,2...24(,25)$, where the $\lambda^i_j$ are the
$25$ $U(5)$ generators. The allowed couplings can then be computed
by tracing over the matrices and using the orthogonality conditions.
Rather than examine every term in the superpotential individually,
we highlight a few of the most important consequences. First, any
term quadratic in the GUT scale fields automatically allows
couplings for every component field, since ${\rm Tr} {\lambda^a
\lambda^b} = 2 \delta^{ab}$ and so for example $(\Phi_1
\tilde{\Phi}_2)= \Phi_1^a \tilde{\Phi}_2^b {\rm Tr} (\lambda^a
\lambda^b) \sim \Phi_1^a \tilde{\Phi}_2^a.$  Second, any term with a
trace containing the VEV of $\Phi_1$ or $\tilde{\Phi}_1$ will
prevent a coupling amongst the $SU(2)$ components; similarly a trace
containing the VEV of $\Phi_2$ or $\tilde{\Phi}_2$ will prevent
$SU(3)$ couplings.  This is due to the zero blocks of these VEVs.
Finally, any term containing a trace over both the VEVs of $\Phi_1$
or $\tilde{\Phi}_1$ and $\Phi_2$ or $\tilde{\Phi}_2$, or a trace
including the VEV of $S$, will be zero. Every mass term will include
either one field of $Z'_4$ R-charge 2 and one of R-charge 0 or two
fields of R-charge 3 or 1 as discussed previously. We will not write
out the results in detail, but they have been confirmed by explicit
computation of the traces and diagonalization of the resulting mass
matrix blocks.  The low energy spectrum of our model indeed
coincides with that of the Pentagon.

\section{Matter Fields}

The purpose of this section is to consider the constraints on the
low energy quark and lepton mass matrices. We have to embed the
standard model fields and the penta-quarks in our model, without
introducing anomalies in either the gauge symmetries or the discrete
$R$ symmetry.

\subsection{R Symmetry Constraints}

In the standard $SU(5)$ GUT theories the chiral matter consists of
three `up' fields ${\cal U}_i, i=1,2,3$ that transform as $10$s,
three `down' fields ${\cal D}_i, i=1,2,3$ transforming as
$\bar{5}$s, and a pair of Higgs with $H_u \in 5$ and $H_d
\in \bar{5}$.  In our model the content will be the same, but we have
some freedom to choose which of the two $SU(5)$s to place these fields
in. In order to cancel chiral anomalies there must be one $\bar{5}$
for each $5$ or $10$ in each $SU(5)$; this allows three possible
configurations (table~\ref{matterconfigs}).

\begin{table}[!h!t!b]
\begin{center}
\noindent \begin{tabular}{|c|c|}
\multicolumn{2}{c} {Configuration 1} \\
\hline
$SU_1(5)$ & $SU_2(5)$ \\
\hline
$H_u = 5$ & $H_d = \bar{5}$ \\
${\cal U}_1 = 10$ & \\
${\cal D}_1 = \bar{5}$ & \\
${\cal D}_2 = \bar{5}$ & ${\cal U}_2 = 10$\\
${\cal U}_3 = 10$ & \\
${\cal D}_3 = \bar{5}$ & \\
\hline
\end{tabular}
 \nolinebreak \hskip 2mm \begin{tabular}{|c|c|}
\multicolumn{2}{c}{Configuration 2} \\
\hline
$SU_1(5)$ & $SU_2(5)$ \\
\hline
$H_u = 5$ & $H_d = \bar{5}$ \\
${\cal U}_1 = 10$ & \\
${\cal D}_1 = \bar{5}$ & \\
${\cal D}_2 = \bar{5}$ & ${\cal U}_2 = 10$\\
& ${\cal U}_3 = 10$\\
& ${\cal D}_3 = \bar{5}$\\
\hline
\end{tabular} \nolinebreak \hskip 2mm \begin{tabular}{|c|c|}
\multicolumn{2}{c}{Configuration 3} \\
\hline
$SU_1(5)$ & $SU_2(5)$ \\
\hline
$H_u = 5$ & $H_d = \bar{5}$ \\
& ${\cal U}_1 = 10$ \\
& ${\cal D}_1 = \bar{5}$ \\
${\cal D}_2 = \bar{5}$ & ${\cal U}_2 = 10$\\
& ${\cal U}_3 = 10$ \\
& ${\cal D}_3 = \bar{5}$ \\
\hline
\end{tabular}
\end{center}
\caption{Anomaly free matter configurations} \label{matterconfigs}
\end{table}

We want to ensure that at least the top quark mass is unsuppressed
at low energies, so we will ignore the third of these possibilities
since it does not allow a renormalizable Yukawa coupling for any of
the up quarks.  In fact, the only cases that will be of interest are
models which allow $H_u {\cal U}_1 {\cal U}_1$ or $H_u {\cal U}_1
{\cal U}_3$ with three generations in $SU_1(5)$ (first
configuration) or $H_u {\cal U}_1 {\cal U}_1$ with the generations
mixed between the two $SU(5)$'s (second configuration).  This requirement
provides the first constraint on the R-charges of the matter fields,
$H_u + {\cal U}_1 + {\cal U}_{1,3} = 2$.

The R-charges of the rest of the fields can be chosen by considering
the desired low-energy mass matrices for the quarks and leptons.
Depending on the choice of matter configuration between the two
$SU(5)$s, a low energy Yukawa coupling will exist only for a given
combination of high energy fields, and this combination must sum to
R-charge 2 for it to appear in the superpotential.  In particular
these terms will contain some number of $\Phi$ or $\tilde{\Phi}$
fields to mediate the interaction between matter fields in separate
$SU(5)$s. The number of such fields is determined by gauge invariance,
but since $\Phi_1,\tilde{\Phi}_1$ and $\Phi_2,\tilde{\Phi}_2$ have
different VEVs the choice between which to include when constructing
the high energy terms will be determined by the desired low energy
content (see tables~\ref{config1},~\ref{config2}).

For instance, the Yukawa coupling $h_u Q_2 \bar{U}_2$ would be
generated by a high energy term containing $H_u, {\cal U}_2, {\cal
U}_2$, and the only gauge invariant construction (that is non-zero
after inserting the $\Phi$ VEVs) is of the form $H_u \tilde{\Phi}_i
{\cal U}_2 {\cal U}_2 \sim ({\rm tr}(5 \times \bar{5}), \det[5 \times 5 \times
5 \times 5 \times 5])$.  In this case the choice of $i$ is clear,
$i=1$ would produce a coupling with the Higgs triplet while $i=2$
would give the desired coupling to the Higgs doublet.  Nevertheless,
this choice has an important consequence: because $\tilde{\Phi}_2$
has R-charge 2, the sum of the matter field R charges must be zero,
not two.

In some cases the choice is not so clear, and in fact the difference
in R-charge between the $\Phi$s can lead to mutually exclusive low
energy couplings.  Consider the coupling $H_u {\cal U}_1 \Phi_i
{\cal U}_2 \Phi_j \sim (\det[5 \times 5 \times 5 \times 5 \times 5],
{\rm tr}[\bar{5} \times 5 \times 5 \times \bar{5}])$.  Evidently the choice
$i=j$ would require the sum of R-charges $H_u + {\cal U}_1 +{\cal
U}_2 = 2$, while $i \neq j$ requires $H_u + {\cal U}_1 +{\cal U}_2 =
0$.  Obviously these conditions cannot both be satisfied, but the
former (with $i=1$) generates the low energy Yukawa coupling $h_u
Q_1 \bar{U}_2$ while the latter leads to $h_u \bar{U}_1 Q_2$.  To
see this, let us represent the ${\cal U}_i$ by the $5 \times 5$
matrix constructed of its low energy components,


$$
\mbox{\Large$\cal U$} = \pmatrix{
& & | & \cr
&  \makebox[40pt]{\LARGE $\overline{U}$} & \vline & \makebox[24pt][l]{ \LARGE $Q$ } \cr
& & | & \cr
\hline
& & | & \cr
& \makebox[40pt]{\LARGE $Q$} & \vline & \makebox[24pt][l]{\LARGE $\overline{E}$} \cr
 }
$$

\noindent In $SU_2(5)$ we will be multiplying $\langle \Phi_i
\rangle {\cal U}_2 \langle \Phi_j \rangle$, so $i=j=1$ selects out
$\bar{U}_2$, $i=j=2$ selects $\bar{E}_2$, and $i \neq j$ gives
$Q_2$.  In $SU_1(5)$ we are taking a determinant of five vectors
each with five components. Let us think of these as column vectors
each with five rows.  We know that the only non-zero contributions
to a determinant will involve the multiplication of components from
unique rows for each vector, {\it i.e.} there must be a component
contribution from each row 1-5.  Now the $\langle \Phi_1 \rangle$
only have non-zero components in rows 1-3 while the $\langle \Phi_2
\rangle$ will only contribute non-zero components from rows 4 and 5.
However, to end up with the Higgs doublet the vector corresponding
to $H_u$ must contribute a component from either row 4 or 5 as well;
thus the determinant including $i=j=2$ will automatically be zero.
If on the other hand $i=j=1$, these will both contribute components
from rows 1-3, so the vectors corresponding to ${\cal U}_1$ must
have one contribution from rows 1-3 and one from rows 4-5, i.e. the
components of $Q_1$.  If $i \neq j$, both contributions from ${\cal
U}_1$ must be in rows 1-3, these components correspond to
$\bar{U}_1$.

In tables~\ref{config1},~\ref{config2} we have listed the high
energy term responsible for each low energy Yukawa coupling as well
as the necessary R-charge sum for the matter fields involved,
dependent on the placement of matter in the two $SU(5)$s.

\begin{table}[!h!t!b]
\begin{tabular}{|c|c|c|}
\hline
High Energy Term & R-charge Requirement & Low Energy Yukawa Couplings \\
\hline
$H_u {\cal U}_{1,3} {\cal U}_{1,3}$ & $H_u+{\cal U}_{1,3} +{\cal U}_{1,3}=2$ & $h_u Q_{1,3} \bar{U}_{1,3}$ \\
$H_u {\cal U}_{1,3} \Phi_1 {\cal U}_2 \Phi_1$ & $H_u + {\cal U}_{1,3} + {\cal U}_2 = 2$ & $h_u Q_{1,3} \bar{U}_{2}$  \\
$H_u {\cal U}_{1,3} \Phi_1 {\cal U}_2 \Phi_2$ & $H_u + {\cal U}_{1,3} + {\cal U}_2 = 0$ & $h_u Q_{2} \bar{U}_{1,3}$ \\
$H_u \tilde{\Phi}_2 {\cal U}_2 {\cal U}_2$ & $H_u + 2 {\cal U}_2 = 0$ & $h_u Q_{2} \bar{U}_{2}$ \\
${\cal D}_{1,2,3} {\cal U}_{1,3} \tilde{\Phi}_2 H_d$ & $H_d + {\cal U}_{1,3} + {\cal D}_{1,2,3} = 0$ & $ h_d Q_{1,3} \bar{D}_{1,2,3}, h_d L_{1,2,3} \bar{E}_{1,3}$ \\
${\cal D}_{1,2,3} \Phi_1 {\cal U}_2 H_d$ & $H_d + {\cal U}_{2} + {\cal D}_{1,2,3} = 2$ & $h_d Q_{2} \bar{D}_{1,2,3}$ \\
${\cal D}_{1,2,3} \Phi_2 {\cal U}_2 H_d$ & $H_d + {\cal U}_{2} + {\cal D}_{1,2,3} = 0$ & $h_d L_{1,2,3} \bar{E}_{2}$ \\
\hline
\end{tabular}
\caption{Yukawa term R-charge constraints (1st configuration)}
\label{config1}
\end{table}

\begin{table}[!h!t!b]
\begin{tabular}{|c|c|c|}
\hline
High Energy Term & R-charge Requirement & Low Energy Yukawa Couplings \\
\hline
$H_u {\cal U}_{1} {\cal U}_{1}$ & $H_u+2 {\cal U}_{1} =2$ & $h_u Q_{1} \bar{U}_{1}$ \\
$H_u {\cal U}_{1} \Phi_1 {\cal U}_{2,3} \Phi_1$ & $H_u + {\cal U}_{1} + {\cal U}_{2,3} = 2$ & $h_u Q_{1} \bar{U}_{2,3}$  \\
$H_u {\cal U}_{1} \Phi_1 {\cal U}_{2,3} \Phi_2$ & $H_u + {\cal U}_{1} + {\cal U}_{2,3} = 0$ & $h_u Q_{2,3} \bar{U}_{1}$ \\
$H_u \tilde{\Phi}_2 {\cal U}_{2,3} {\cal U}_{2,3}$ & $H_u + {\cal U}_{2,3} + {\cal U}_{2,3} = 0$ & $h_u Q_{2,3} \bar{U}_{2,3}$ \\
${\cal D}_{1,2} \Phi_1 {\cal U}_{2,3} H_d$ & $H_d + {\cal U}_{2,3} + {\cal D}_{1,2} = 2$ & $ h_d Q_{2,3} \bar{D}_{1,2}$ \\
${\cal D}_{1,2} \Phi_2 {\cal U}_{2,3} H_d$ & $H_d + {\cal U}_{2,3} + {\cal D}_{1,2} = 0$ & $ h_d L_{1,2} \bar{E}_{2,3}$ \\
$H_d \tilde{\Phi}_2 {\cal U}_1 \tilde{\Phi}_1 {\cal D}_3$ & $H_d + {\cal U}_1 + {\cal D}_3 = 0$ & $h_d Q_1 \bar{D}_3$ \\
$H_d \tilde{\Phi}_2 {\cal U}_1 \tilde{\Phi}_2 {\cal D}_3$ & $H_d + {\cal U}_1 + {\cal D}_3 = 2$ & $h_d L_3 \bar{E}_1$ \\
${\cal D}_{1,2} {\cal U}_1 \Phi_2 H_d$ & $H_d + {\cal U}_{1} + {\cal D}_{1,2} = 0$ & $h_d Q_1 \bar{D}_{1,2}, h_d L_{1,2} \bar{E}_{1}$ \\
\hline
\end{tabular}
\caption{Yukawa term R-charge constraints (2nd configuration)}
\label{config2}
\end{table}

Another constraint on the R-charges is that the 3 low energy
operators $h_u L_{1,2,3}$ are forbidden.  These would give a GUT
scale mass to one of the Higgs fields and a matter field,
eliminating them from the low-energy spectrum.

Finally, there is also a constraint the R-charges must satisfy in
order to ensure that the discrete $Z_4$ R-symmetry is anomaly free.
This gives the equations (mod 4):

$$
\begin{array}{lll}
SU_1(5): & 0 = & 10 \lambda + 5 (\Phi_1 + \tilde{\Phi}_1 + \Phi_2 + \tilde{\Phi}_2 -4) + 10 p ( T - 1) +10 (S-1) \\
& & + 5 (P + \tilde{P} -2)  + (H_u -1) + \sum_i 3({\cal U}_i -1) + \sum_j ({\cal D}_j -1) \\
& &\\
SU_2(5): & 0 =& 10 \lambda + 5 (\Phi_1 + \tilde{\Phi}_1 + \Phi_2 + \tilde{\Phi}_2 -4) + 10 (1-p) (T - 1) \\
& & + (H_d -1) + \sum_i 3({\cal U}_k -1) + \sum_j ({\cal D}_l -1)
\end{array}.
$$

The $\lambda$s represent the gauginos, which must have R-charge 1
(the vector fields having R-charge 0) since they arise in the D-term
of the superpotential.  $p = 0,1$ indicating which $SU(5)$
the $T$ transforms in (notice that we could have created a similar
parameter for the $S$ and $P,\tilde P$, but their contributions will
sum to zero anyway).  $i,j$ run over the matter fields in
$SU_1(5)$, and $k,l$ $SU_2(5)$.  Inserting the known R-charge values for
the fields, these simplify to:

$$
\begin{array}{l}
SU_1(5): 1 + 2 p + H_u + \sum_i 3({\cal U}_i -1) + \sum_j ({\cal D}_j -1) = 0 \\
SU_2(5): 1 + 2 (1-p) + H_d + \sum_k 3({\cal U}_k -1) + \sum_l
({\cal D}_l -1) = 0
\end{array}.
$$

\noindent These constraints allow us to determine the R-charge of
two of the matter fields in terms of the others (recall that $H_u$
can be determined by the requirement of a renormalizable top quark
Yukawa term, and $H_d$ can be determined by $\tilde{\Phi}_1 + H_u +
H_d = 2$).  On the other hand, we can use the last of these conditions to combine the two equations into a single constraint on the matter fields

$$
2=3({\cal U}_1+{\cal U}_2+{\cal U}_3) + ({\cal D}_1+{\cal D}_2+{\cal D}_3).
$$

Hence, there is an anomaly-free model for any choice of R-charges which satisfy this condition on the matter fields. However, the R-charges of the Higgs
fields are then uniquely determined from the matter fields.

At this point it should be clear that, although our model has the
field content of the Pentagon model, the low energy R charge
assignments must be quite different.  In particular, we find that
{\it the low energy R charges cannot be generation blind}.  This is
a consequence of the GUT structure, and the anomaly constraints.
There is some freedom to shift the discrete anomaly constraints by
assigning the axion superfield $X$ an additive transformation law
under $Z_4$.   However, as we will see in the next section, this
cannot solve the most severe phenomenological problems of this model.

\subsection{Phenomenology}

An important phenomenological constraint on any GUT is that it
satisfy the experimental bounds on proton decay.  The lower bound on
the overall lifetime of the proton is currently $2.1\times10^{29}$
years \cite{Ahmed:2003sy}.  However, there are stronger bounds for
specific decays, the strongest of which is $1.6\times 10^{33}$ years
for $p\rightarrow e^+ \pi$ \cite{Shiozawa:1998si}.  The triplet
Higgs is no danger to proton decay in this model since it naturally
acquires a GUT-scale mass. However, depending on the choice of
R-charges for the matter fields, there are a number of potentially
dangerous baryon and lepton violating operators that could mediate
proton decay. Dimension-6 operators are suppressed by ${1\over
M_U^2}$, so the decay rates are suppressed roughly by $({m_p\over
M_U})^4 \approx 10^{-64}$.\footnote{Recall that in the Pentagon
model, the unified coupling is on the edge of the perturbative
regime, so there is no significant coupling constant suppression of
proton decay rates.}  This is right in the neighborhood of the
current bound, meaning that it is not ruled out yet but predicts
that proton decay should be seen soon. However, dimension-4 and
dimension-5 operators which violate baryon and lepton number should
not be allowed as they involve fewer inverse-powers of $M_U$ and
would permit proton decay at a rate far outside of current bounds.
An exception to this is dimension-5 {\it purely baryon} number
violating operators, and dimension-5 {\it purely lepton} number
violating operators.  In these cases, two vertices each with an
inverse-power of $M_U$ are required in the same diagram, making the
overall lifetime on the same order as what a dimension-6 operator
which violates {\it both} baryon and lepton number would yield.
Table~\ref{BandL} enumerates the dangerous dimension 4 and 5
operators that could appear in the theory.

\begin{table}[!h!t!b]
\begin{center}
\begin{tabular}{|c|c|c|}
\hline
operator & violation & dimension \\
\hline
$LL\bar{E}$        & L & 4 \\
$LQ\bar{D}$        & L & 4 \\
$\bar{D}\bar{D}\bar{U}$ & B & 4 \\
$\bar{U}\bar{E}D$            & BL & 5 \\
$LQQQ$                       & BL & 5 \\
$\bar{E}\bar{U}\bar{U}\bar{D}$  & BL & 5 \\
\hline
not dangerous: \\
\hline
$Q\bar{U}\bar{L}$  & L & 5 \\
$h_dQQQ$                & B & 5 \\
$h_u h_u L L$ & L & 5 \\
\hline
\end{tabular}
\end{center}
\caption{B and L violating operators} \label{BandL}
\end{table}

We have determined that it is not possible to forbid all of the
dangerous operators with any combination of $Z_4$ R-charges, even
ignoring the discrete anomaly constraints.  Most of
these can be forbidden if we impose matter parity--a discrete (non-R)
$Z_2$ symmetry where all of the matter fields have charge 1 and all
other fields remain
uncharged.  Then, all but the ${\cal UUUD}$ operators are eliminated,
while not forbidding any of the Higgs Yukawa couplings ($H_u {\cal UU}$ or
$H_d{\cal UD}$). This leaves less
work for the R symmetry to do.\footnote{The $Z_2$ matter parity is optimal in the
sense that expanding it to any larger symmetry cannot help forbid
the ${\cal UUUD}$ operators without also forbidding either some of
the Higgs Yukawa couplings or $\tilde{\Phi}_1 H_u H_d$.  The reason
for this is that $\tilde{\Phi}_1$ cannot get a charge under this new
symmetry without forbidding some of the low energy Yukawa couplings
or allowing some of the B or L violating operators by giving T a
charge.  And without a charge for $\Phi_1$, the condition that
$\Phi_1 H_u H_d$, $H_u {\cal UU}$, and $H_d{\cal UD}$ be chargeless
forces ${\cal UUUD}$ to be chargeless.}  After adding the $Z_2$, it
is then possible to forbid all the remaining dangerous operators
with the $Z_4$, however all of those models have a $Z_4$ anomaly
(and the $Z_2$ is anomalous as well).  Later, we will discuss possible
ways of fixing these anomaly problems.  Another possibility is to add
a discrete symmetry (either instead of the $Z_2$ or in addition to it)
which gets rid of ${\cal UUUD}$ but forbids some of the Higgs Yukawa
couplings, which may or may not already be forbidden by the R-symmetry.

We do want to allow the seesaw operators $h_uh_uLL$.  These give a
tiny Majorana mass matrix for the neutrinos. Therefore, we wish to
allow this operator for as many combinations of L generations as
possible.

Another phenomenological issue is that of neutron-anti-neutron
oscillations. This is similar to proton decay in that dimension-6
operators are okay whereas dimension-4 and dimension-5 operators are
not.  Only baryon number violation is relevant for neutron
oscillations, however $n \rightarrow \bar{n}$ requires the baryon
number change by 2.  So the dimension-5 baryon number violating
operator $h_d QQQ$ listed as ``not dangerous'' for proton decay is
also safe for neutron oscillations because it violates baryon number
only by 1.  Hence, nothing new is added by this constraint.

Next we examine the constraints on quark and lepton mass matrices.
The generation dependence of R charges implies that many of the
entries in these mass matrices are zero. Which ones are non-zero,
how many non-zero mass eigenvalues, and the approximate ratios of
mass eigenvalues depends on the choice of R-charges for the matter
fields.  We have used a brute-force computer algorithm to explore
the various possibilities. The first limitation we have found is
that, even ignoring B or L violation as well as the discrete anomaly
constraints, it is impossible to get rank 3 mass matrices for both
the up quarks and the down quarks at the same time.  The requirement
that the $SU(5)$ anomalies cancel implies that there can be at most
5 massive quarks. Models with 5 massive quarks limit the number of
massive neutrinos to 2.  If we choose the up quark to be massless,
then the number of massive leptons is also limited to 2. If instead
we choose a massless down quark, then all 3 leptons can be massive.
Note that in this analysis we have not imposed the constraint that
the B and L violating operators be forbidden--when these constraints
are combined, the restrictions are much more severe.

If we impose the anomaly constraints, there are 512 different models with an unsuppressed top quark mass
in the first configuration of Table~\ref{matterconfigs} and 896 models
in the second configuration.  We have found candidates where a discrete
symmetry other than matter parity is used to forbid some of the dangerous
operators. One possibility is to use a $Z_2'$ symmetry where all of
the ${\cal U}$ fields are odd, and all the rest of the fields are even.
This forbids all of the ${\cal H}_d{\cal U D}$ Yukawa couplings, but still allows
$H_u {\cal U U}$.  It forbids all of the dangerous operators except
${\cal UUD}$, which can either be forbidden by the other $Z_2$ or by
the $Z_4$ R-symmetry. There are a number of models where it is forbidden
by the $Z_4$ alone, 16 of which look potentially interesting (see appendix, Table A.1).  8 of
these involve just a mass for the top quark, all the other quarks
and leptons massless, and either 2 or 3 neutrinos.  The other 8 involve
both top and charm masses, the other quarks and leptons massless, and 2 or 3
neutrinos.  In half of those 8, the charm quark is suppressed by a factor
of $\epsilon^2$ relative to the top quark, which appears preferable
to the other half in which it's suppressed by just a single power of
$\epsilon$.  If instead of just relying on $Z_2'$ (and the $Z_4$ R
symmetry) both $Z_2$ and the $Z_2'$ are imposed, there are no restrictions
on the R charges and there are many more possibilities for the up-quark,
lepton, and neutrino masses.  The main limitation with any models involving
the $Z_2'$ is that all 3 generations of down-type quarks and leptons must
remain massless.

There are other candidate models which involve adding discrete
horizontal symmetries (see appendix, Table A.4). In contrast to adding discrete symmetries
that act the same way on each generation, it takes more creativity
to find horizontal symmetries that work phenomenologically and we
have not yet found a way of automating the process.  Our approach
so far has been to look at models with only 1 ${\cal UUUD}$ operator (since
all the rest can be forbidden by the $Z_2$), and to use a horizontal
symmetry to eliminate this operator.  The symmetry will almost always
remove some Yukawa couplings, but for the case of only a single
${\cal UUUD}$ operator, it can be chosen carefully so as not to
reduce the rank of the mass matrix.  An explicit example which employs a $Z_{3H}$ horizontal symmetry is shown in Table~\ref{horizontal}.
The $Z_{3H}$ symmetry disallows the ${\cal U}_3$ field from coupling to
any other matter fields except as a cubic.  This has the effect of
eliminating the remaining dangerous operator, but also removes the
mass for the charm quark (which in this particular case is good because
it would have been on the order of the top mass).  Furthermore, unlike
the $Z_2$ and $Z_2'$, this $Z_{3H}$ symmetry is itself anomaly free (both
the $Z_2$ and $Z_2'$ have an odd number of charged fields in each $SU(5)$,
whereas the only field charged under the $Z_{3H}$ in this model is a
$ {\cal U}$ which contains a factor of three in the anomaly equation).
The end result is a model with a heavy top quark, a bottom and tau quark
with suppressed masses, and 2 neutrino masses.  Without violating anomalies
or adding more fields, models of this type are the closest to
the real world that we have been able to find.  A slight variation on
this model with the same R-charges is to use a $Z_{4H}$ horizontal symmetry
instead of the $Z_{3H}$ (see appendix, Table A.3).  This results in a model
with a heavy top quark, a bottom and tau quark suppressed by $\epsilon$, and
a charm quark suppressed by $\epsilon^2$.  However, the $Z_{4H}$ symmetry is
itself anomalous.

\begin{table}[!h!t!b]
\begin{center}
\begin{tabular}{|c|c|c|c|}
\hline
 field & R & $Z_2$ & $Z_{3H}$  \\
\hline
 $H_u$ &2 & 0& 0 \\
  $H_d$ &0 &0 &0  \\
 ${\cal U}_{1}$ & 0 &1 & 0  \\
 ${\cal U}_{2}$ & 0 &1 & 0 \\
 ${\cal U}_{3}$ & 0 &1 & 1 \\
 ${\cal D}_{1}$ & 1 &1 & 0 \\
 ${\cal D}_{2}$ & 1 &1 & 0 \\
 ${\cal D}_{3}$ & 0 &1 & 0 \\
\hline
\end{tabular}
\end{center}
\caption{Example of a model with $Z_{3H}$ horizontal symmetry (configuration 2) in addition to matter parity. Model contains a heavy top, a lighter bottom and tau, two neutrinos, and the other two generations of quarks and leptons massless.}
\label{horizontal}
\end{table}

We have found that shifting the discrete anomaly equations by
giving the $X$ field a transformation law under the $Z_4$, does not help us
to obtain full rank mass matrices. It also does not help fix the anomaly in
the $Z_2$ or $Z_{4H}$ symmetries, although it can be used to fix the $Z_2'$
anomaly.  If these models are to be made
realistic, one would have to imagine that the missing matrix
elements of the quark and lepton mass matrices came from breaking of
the R symmetry. In CSB explicit R symmetry breaking is expected to
vanish like a power of the cosmological constant.   The gravitino
mass is $\sim \Lambda^{1/4}$.   In the dimensionless quark and
lepton Yukawa couplings we might imagine $\Lambda^{1/4} / m$. Even
if we take $m$ of order a TeV, this is too small to be of
phenomenological help.  We would have to postulate corrections that
scale with even smaller powers of $\Lambda$.   Furthermore, since we
have used the R symmetry to eliminate dangerous B and L violating
operators, one would have to explain why these R violating
corrections did not lead to rapid proton decay.

\section{ {Adding a {$10$} and a {$\overline{10}$} }}

An interesting possibility for resolving the issue of full rank mass
matrices is to add an extra pair of Higgs fields.  As discussed in
the previous section, because the $SU(5)$ anomalies force us to place
the matter fields in different gauge groups (and in particular forces us to split one of the generations), certain Yukawa terms
are formed only by inserting VEVs of the $\Phi$ fields to bridge the
two $SU(5)$s, and this prevents certain low energy couplings due to
R-charge.  However, by inserting an extra pair of $\bar{5},5$ fields
into the two $SU(5)$s, we can place all of the matter in a single
$SU(5)$. As before we will want to ensure that all Higgs triplets gain
a GUT scale mass, so we will enforce that $H_u + H_d =2$, as well as
$G_d + G_u =2$ where $G_d$ is a $\bar{5}$ in $SU_1(5)$ and
$G_u$ a $5$ in $SU_2(5)$. We will also impose that $H_u+G_d
\neq 2$ and $H_d+G_u \neq 2$ since these could give mass to the
doublets.  Then, rewriting the $Z_4$ anomaly constraints,

$$
\begin{array}{l}
SU_1(5):0=  2 p + H_u + G_d + 3({\cal U}_1+{\cal U}_2+{\cal U}_3) + ({\cal D}_1+{\cal D}_2+{\cal D}_3)\\
SU_2(5):0= 2 (1-p) + H_d + G_u
\end{array},
$$

\noindent and imposing the conditions asserted above, we find that
$p = 1$, $H_d = -G_u, H_u = -G_d$, and so $2=3({\cal
U}_1+{\cal U}_2+{\cal U}_3) + ({\cal D}_1+{\cal D}_2+{\cal D}_3)$.
That is, the R-charges of the Higgs fields are determined in terms
of one another, but are completely independent of the charges of the
matter fields.

There is, unfortunately, a significant phenomenological problem with
this idea: in the model above all of the triplet Higgs do gain mass,
but this means all of the doublets remain massless.  Introducing a
new pair of low energy Higgs doublets introduces eight new degrees
of freedom, four of which are new charged bosons, which would
probably be a disaster in terms of flavor changing neutral currents.
In fact this problem is unavoidable, even if we relax the anomaly
constraints imposed above. In order to ensure that all triplet Higgs
gain mass, we must have $H_u + H_d =2$ and $G_d + G_u =2$, or $H_u +
H_d+G_d + G_u =0$. This implies that at low energies $h_u+h_d = -g_u
- g_d$.  However, we want our original pair of Higgs doublets to
have R-charge such that $h_u + h_d = 0$ to ensure that they remain
massless at the TeV scale as well as to allow a $\mu$ term $s h_u
h_d$; therefore {\it both} Higgs doublets must remain massless as
long as we insist on avoiding light triplets.  Possible resolutions
of this issue are discussed in \cite{decon}.

This does lead us to a slightly different approach, however. Instead
of an extra pair of Higgs fields, let us introduce two fields $A
\in (\overline{10},1)$ and $B \in (1,10)$ such that $A+B=2$. These
fields will have zero VEVs so will not affect our SUSY vacuum, and
they will gain GUT scale mass because both the terms $A \Phi_1
\Phi_1 B$ and $A \Phi_2 \Phi_2 B$ are allowed (actually this is a
total of four terms since there are two different ways to contract
the indices). The important point is that, as in the model with two
pairs of Higgs doublets, all of the matter fields can be placed in a
single $SU(5)$.  The $Z_4$ anomaly constraints are now

$$
\begin{array}{l}
SU_1(5):0= 2 + 2 p + (H_u-1) + 3(A-1) + 3({\cal U}_1+{\cal U}_2+{\cal U}_3-3) + ({\cal D}_1+{\cal D}_2+{\cal D}_3-3)\\
SU_2(5):0= 2 + 2 (1-p) + (H_d-1) + 3(B-1)
\end{array}.
$$

\noindent The second of these defines a relation between the Higgs and the $10, \overline{10}$ fields.  Combining this with the requirements that $H_u + H_d =2$ and $A+B=2$, we now find that $H_u$ and $A$ cancel each other in the first equation, and so we are left with

\begin{eqnarray}
A &=& H_u+2 p \\
2 &=& 3({\cal U}_1+{\cal U}_2+{\cal U}_3) + ({\cal D}_1+{\cal D}_2+{\cal D}_3)
\label{eqn:matteranomaly}
\end{eqnarray}

\noindent just as for the case of two Higgs doublet pairs above.
Again, the R-charges of the $H_u,H_d,A,B$ are all related but independent of the charges of the matter fields.  Remarkably, the relationship between the R-charges of the matter fields turns out to be the same regardless of whether we add the extra pair of Higgs, the A and B fields, or neither.  The results will be discussed below.

First let us note that while this discussion has so far been specific to models with all matter in a single $SU(5)$, the inclusion of $A$ and $B$ actually provides more freedom for the placement of the fields.  In fact, all that is required by gauge anomaly cancellation is that each {\it generation} of matter be placed in a single $SU(5)$ (Table ~\ref{matterconfigsAB}).  We will ignore configuration 4 because the top quark mass is suppressed.   Configuration 1 has been discussed so far.  More generally, for any of the configurations, the anomaly equations can be put in the form:

$$
\begin{array}{l}
SU_1(5): 2 + 2 p + H_u + 3 A + \sum_i 3({\cal U}_i -1) + \sum_j ({\cal D}_j -1) = 0 \\
SU_2(5): 2 + 2 (1-p) + H_d + 3 B + \sum_k 3({\cal U}_k -1) + \sum_l ({\cal D}_l -1) = 0
\end{array}.
$$

\noindent These equations can be combined to give exactly the same relation between the matter fields as above (\ref{eqn:matteranomaly}).  In our models without A and B, the Higgs R-charges were then determined in terms of the matter fields.  However, in the present case, the Higgs R-charges can also be freely chosen (as long as they add up to 2) and it's A and B which are uniquely determined.  This would appear to give rise to four times as many anomaly-free distinguishable low energy models.  This is exciting, however in actuality it only adds twice as many new models, because of some redundancy due to the p parameter.  Another advantage is that, in all of these configurations, our $Z_2$ matter parity is anomaly free (due to there being an even number of fields in each $SU(5)$).

\begin{table}[!h!t!b]
\begin{center}
\noindent \begin{tabular}{|c|c|}
\multicolumn{2}{c} {Configuration 1} \\
\hline
$SU_1(5)$ & $SU_2(5)$ \\
\hline
$H_u = 5$ & $H_d = \bar{5}$ \\
$A = \overline{10}$ & $B = 10$ \\
${\cal U}_1 = 10$ & \\
${\cal D}_1 = \bar{5}$ & \\
${\cal U}_2 = 10$ & \\
${\cal D}_2 = \bar{5}$ & \\
${\cal U}_3 = 10$ & \\
${\cal D}_3 = \bar{5}$ & \\
\hline
\end{tabular}
 \nolinebreak \hskip 2mm \begin{tabular}{|c|c|}
\multicolumn{2}{c}{Configuration 2} \\
\hline
$SU_1(5)$ & $SU_2(5)$ \\
\hline
$H_u = 5$ & $H_d = \bar{5}$ \\
$A = \overline{10}$ & $B = 10$ \\
${\cal U}_1 = 10$ & \\
${\cal D}_1 = \bar{5}$ & \\
 ${\cal U}_2 = 10$ & \\
${\cal D}_2 = \bar{5}$ &\\
& ${\cal U}_3 = 10$\\
& ${\cal D}_3 = \bar{5}$\\
\hline
\end{tabular} \nolinebreak \hskip 2mm \begin{tabular}{|c|c|}
\multicolumn{2}{c}{Configuration 3} \\
\hline
$SU_1(5)$ & $SU_2(5)$ \\
\hline
$H_u = 5$ & $H_d = \bar{5}$ \\
$A = \overline{10}$ & $B = 10$ \\
& ${\cal U}_1 = 10$ \\
& ${\cal D}_1 = \bar{5}$ \\
& ${\cal U}_2 = 10$\\
& ${\cal D}_2 = \bar{5}$ \\
 ${\cal U}_3 = 10$ & \\
 ${\cal D}_3 = \bar{5}$ & \\
\hline
\end{tabular}
\hskip 2mm \begin{tabular}{|c|c|}
\multicolumn{2}{c}{Configuration 4} \\
\hline
$SU_1(5)$ & $SU_2(5)$ \\
\hline
$H_u = 5$ & $H_d = \bar{5}$ \\
$A = \overline{10}$ & $B = 10$ \\
& ${\cal U}_1 = 10$ \\
& ${\cal D}_1 = \bar{5}$ \\
& ${\cal U}_2 = 10$ \\
& ${\cal D}_2 = \bar{5}$ \\
& ${\cal U}_3 = 10$ \\
& ${\cal D}_3 = \bar{5}$ \\
\hline
\end{tabular}
\end{center}
\caption{Anomaly free matter configurations with a 10 and $\overline{10}$} \label{matterconfigsAB}
\end{table}

Our most significant result is the discovery of models which are free of all B and L violation and have entirely full mass matrices.  Unfortunately, these models suffer from an anomaly in the R-symmetry, but we believe this can be remedied with the axion.  Without the axion, we find that it is still not possible to form $Z_4$ anomaly-free models with full rank mass matrices, regardless of the matter configuration.  It is possible to form anomaly-free models with everything massive except the up quark.  However, even these models are problematic, as they all allow ${\cal U} {\cal U} {\cal U} {\cal D}$  B and L violating operators; as discussed previously, these operators cannot be
eliminated with generation blind discrete symmetries.  Nevertheless, these models were not even possible before the inclusion of the $A$ and $B$ fields.  The bottom line is that, by allowing each matter generation to sit in a single $SU(5)$, we have introduced a plethora of new models, both with and without anomalies, that appear much more realistic than what was previously possible.  
The results of our computer search can be summarized as follows:
\begin{itemize}
\item There is a class of anomaly-free models in configurations 1,3 that have masses for 2
up quarks, 3 down quarks, 3 leptons, and 2 or 3 neutrinos; however,
there are ${\cal U} {\cal U} {\cal U} {\cal D}$ B and L violating
operators that need to be removed with a horizontal symmetry (see appendix, Table A.8).

\item There are $Z_4$ anomaly-free models from all three configurations that have non-trivial mass matrices and no ${\cal UUUD}$ operators.  The remaining B and L violating operators can be forbidden by our (anomaly-free) $Z_2$ (see appendix, Table A.9).

\item All three configurations have anomaly-free models without dangerous baryon violating operators, and only 3 (configuration 1) or 9 (configurations 2,3) lepton violating operators.  Most of these have all quarks and leptons massless, but there are a few in configurations 2,3 with 2 up quarks, 1 down quark, 0 leptons, and 2 neutrinos.  See appendix, Table A.10.

\item There are no anomaly-free models that are also free of lepton violation, but a large class of them (with and without baryon violation) if we ignore the anomaly constraint.  These might be
useful for resolving puzzles associated with the current experimental bound on the Higgs mass.

\item There are very interesting models in configuration 1 (all matter in $SU_1(5)$) that have a $Z_4$ anomaly but there are NO baryon and lepton violating operators and ALL
Yukawa couplings (every entry in each matrix) are non-zero (Table ~\ref{ideal}).  These
models have a possible Froggatt-Nielsen mechanism.  Certain Yukawa
couplings involve one higher power of the GUT scale fields $\Phi$,
so if the VEV is $\epsilon$ in units of $M_U$, some matrix elements
will be suppressed.  One finds no suppression for up quarks or for
the neutrino seesaw terms and one power of $\epsilon$ for downquarks
and leptons.  This could supply part of the explanation of the
texture of mass matrices.   For the rest we would have to invent
more Froggatt-Nielsen symmetries.
\end{itemize}

\begin{table}[!h!t!b]
\begin{changemargin}{-0.5in}{-0.5in}
\begin{center}
\begin{tabular}{|c|ccccc|cccccc|cc|cccc|}
\hline
config. & $H_u$ & $H_d$ & $p$ & $A$ & $B$ & ${\cal U}_{1}$  & ${\cal U}_{2}$ & ${\cal U}_{3}$ & ${\cal D}_{1}$ & ${\cal D}_{2}$ & ${\cal D}_{3}$ & B & L  & Ups & Downs & Lept. & Neut.   \\
\hline
  1st &0 &2 & 0 & 0 &2 &1 &1 &1 &1 &1 &1 &0 &0  &3 &3 &3 &3 \\
  1st  &0 &2 &0 & 0 &2 &3 &3 &3 &3 &3 &3 &0 &0  &3 &3 &3 &3 \\
1st &0 &2 &1 &2 &0 &1 &1 &1 &1 &1 &1 &0 &0  &3 &3 &3 &3 \\
  1st &0 &2 &1 & 2 &0 &3 &3 &3 &3 &3 &3 &0 &0  &3 &3 &3 &3 \\
\hline
\end{tabular}
\end{center}
\end{changemargin}
\begin{changemargin}{-0.25in}{-0.25in}
\caption{Ideal models: no B or L violation and full quark and lepton mass matrices.  All the up quarks are unsuppressed, downquarks and leptons are suppressed by $\epsilon$, and neutrinos are unsuppressed.  These models are ruled out by the instanton anomaly but are made possible by adding an axion.}
\label{ideal}
\end{changemargin}
\end{table}

The models of Table ~\ref{ideal} can be rendered anomaly free by
assigning a $Z_4$ transformation law to the axion. In these models
all matter fields have identical R-charge (= 1 or 3), and are precisely
 the charge assignments in the original Pentagon papers.
Specifically, for all matter fields to have the same R-charge we
want $3({\cal U}_1+{\cal U}_2+{\cal U}_3) + ({\cal D}_1+{\cal
D}_2+{\cal D}_3)=0$, instead of $2$. This can be arranged with an axion
shift of $\pi$ in units of $f_a$ in $SU_1(5)$.  These models are
ideal in a number of ways and stand out as clearly superior to all
other models discussed in this paper.

Since the axion decay constant is large, one might have worried that
this mechanism will lead to large spontaneous breaking of the $Z_4$
symmetry. In other words, we can replace non-invariant operators by
invariant ones, simply by multiplying with the appropriate power of
$e^{X/f_a}$.  This is mathematically correct, however, if $X$ is
really to serve as a QCD axion, no such terms can appear in the
effective action above the QCD scale.  If they did, they would
provide a potential for the axion which dominates that generated by
QCD, and $X$ would not solve the strong CP problem. In our model, we
include $X$ in the GUT scale Lagrangian only via a term $$\int d^2
\theta\ (X/f_a)\ W_{\alpha}^2 + h.c.$$ involving the gauge field
strength of $SU_1(5)$.   The classical Lagrangian has a $U(1)$ shift
symmetry, which is preserved to all orders in perturbation theory,
and broken predominantly by QCD\footnote{Here we are assuming that
SUSY is broken by the ISS mass term.  In the supersymmetric limit of
the Pentagon model, QCD is IR free and no potential is generated for
$X$.}. One has to appeal to a more fundamental UV complete model, to
justify the argument that there are no other couplings of $X$
allowed in the GUT scale effective Lagrangian.

\section{Conclusions}

In this paper, we have examined a large variety of SUSic grand
unified models, which solve the doublet triplet splitting problem
and reduce to the Pentagon model at low energies.  Most of the
models have phenomenological problems, and all of them have a large
number of GUT scale fields. We are not particularly bothered by the
latter problem because we view our models as a stepping stone to
higher dimensional models originating in string theory.

The most successful class of models involved the addition of GUT
scale fields $A$ and $B$, transforming in the $(\overline{10}, 1) \oplus
(1,10)$ of $SU_1(5) \times SU_2(5)$. If we use the QCD axion to cancel
the discrete anomaly, we
obtain models which preserve Baryon and Lepton number up to and
including dimension $5$ operators.  They also have quark and lepton
mass matrices of full rank.  The models {\it could } predict a
hierarchy between up quark masses and those of down quarks and
charged leptons\footnote{Recall that in the Pentagon model we expect
that $\tan\beta \sim 1$, so that the up-down hierarchy must be
attributed to Yukawa couplings.}, if a certain VEV is small in GUT
units.  It is not clear whether it makes sense to attribute the
entire ratio $m_t / m_b$ to a single power of the VEV of a scalar
field, and one would have to understand more about the microscopic
origin of the model before claiming this as a victory.

The rest of the texture of the quark and lepton mass matrices might
be explainable in terms of horizontal symmetries and the
Froggatt-Nielsen mechanism.  In models without the $A$ and $B$
fields, we have been forced to introduce discrete symmetries to
eliminate dangerous B and L violating operators.  For the most part,
these led to unpleasant results for mass matrices, with too many
massless particles. As far as we can see, the only viable strategy
for such models is to postulate terms which explicitly break the
$Z_4$ symmetry, above and beyond the ISS mass term.  In order to
get acceptable results, we would probably have to postulate R
breaking terms that scale to zero even more slowly than the (already
mysterious) $\Lambda^{1/4}$.  Of course, if we abandon the origins
of the Pentagon model in CSB, and attribute the R breaking to
dynamical mechanisms in effective field theory, the range of
possibilities is wider. We have not explored this option.

Our preference is to pursue the addition of horizontal symmetries to
the models with $A$ and $B$ fields, which have full rank mass
matrices in their current form. This will be the subject of future
work. Another direction we want to pursue is a loosening of our
requirements on dimension four $B$ and $L$ violation.  This might be
useful for resolving puzzles associated with the current
experimental bound on the Higgs mass.

\section{Acknowledgments}

We would like to thank M.Dine for discussions.  This research was
supported in part by DOE grant number DE-FG03-92ER40689.

\begin{changemargin}{-0.5in}{-0.5in}

\newpage

\section{Appendix}

In the appendix we list for reference a variety of classes of models that we find interesting.  The list is not comprehensive, but contains models we believe may be useful in a future search for a Froggatt-Nielson mechanism.   For each model we list its $SU(5)$ matter configuration (Tables ~\ref{matterconfigs},~\ref{matterconfigsAB} in the text), the R-charges of the relevant fields, the number of B (baryon) and L (lepton) violating operators allowed by the R-symmetry (before imposing other discrete symmetries), the number of ${\cal U} {\cal U} {\cal U} {\cal D}$ operators where relevant, and the ranks of the up quark, down quark, lepton, and neutrino mass matrices.

\subsection{Models without a $10$ and $\overline{10}$}

\begin{center}
\begin{tabular}{|c|ccc|cccccc|cc|cccc|}
\multicolumn{16}{c}{Table A.1} \\
\hline
config & $H_u$ & $H_d$ & $p$ & ${\cal U}_{1}$  & ${\cal U}_{2}$ & ${\cal U}_{3}$ & ${\cal D}_{1}$ & ${\cal D}_{2}$ & ${\cal D}_{3}$ & B& L & Ups & Downs & Lept. & Neut.   \\
\hline
 1st& 0& 2& 0& 2 &2 &1 &1 &1 &1 &30 &33 & 2 &0 &0 &3 \\
 1st& 2& 0& 1& 2& 2& 1& 1& 1& 1&30 &33 &1 &0 &0 &3 \\
 1st& 0& 2& 0& 1 &2 &2 &1 &1 &1 &30 &33 & 2 &0 &0 &3 \\
 1st& 2& 0& 1& 1& 2& 2& 1& 1& 1&30 &33 &1 &0 &0 &3 \\
 1st& 0& 2& 0& 3& 2& 2& 3& 3& 3&30 &33 &2 &0 &0 & 3\\
 1st& 2& 0& 1& 3& 2& 2& 3& 3& 3&30 &33 &1 &0 &0 & 3\\
 1st& 0& 2& 0& 2 & 2 & 3  &3 &3 &3 &30 &33 &2 &0 &0 &3 \\
 1st& 2& 0& 1& 2 & 2&3 &3 &3 &3 & 30 & 33 &1 &0 &0 &3\\
 2nd & 2& 0& 1 & 0 & 0 &0 &1 &3 &2 &18 &27 &1 &0 &0 &2 \\
 2nd & 2& 0& 1 & 0 & 0 &0 &3 &1 &2 &18 &27 &1 &0 &0 &2 \\
 2nd & 2& 0& 1 & 2 & 2 &2 &1 &1 &2 & 1 & 7 &1 &0 &0 &2 \\
 2nd & 2& 0& 1 & 2 & 2 &2 &3 &3 &2 & 1 & 7 &1 &0 &0 &2 \\
 2nd & 2& 0& 1 & 2 & 2 &1 &1 &1 &1 &32 &47 &2 &0 &0 &3 \\
 2nd & 2& 0& 1 & 2 & 1 &2 &1 &1 &1 &32 &47 &2 &0 &0 &3 \\
 2nd & 2& 0& 1 & 2 & 3 &2 &3 &3 &3 &32 &47 &2 &0 &0 &3 \\
 2nd & 2& 0& 1 & 2 & 2 &3 &3 &3 &3 &32 &47 &2 &0 &0 &3 \\
\hline
\end{tabular}
\end{center}

Table A.1 shows 16 models with masses for either a top and charm or just a top and no down-type quarks or lepton masses.  The charm is suppressed by $\epsilon$ for models in configuration 1 and by $\epsilon^2$ for models in configuration 2.  All B,L violating operators are forbidden by imposing a $Z_2'$ with ${\cal U} =1 $ , ${\cal D} = 0$, which also eliminates all $H_d {\cal U} {\cal D}$ (note: the number of B and L violating operators listed are before applying this symmetry).

\begin{center}
\begin{tabular}{|c|c|c|c|}
\multicolumn{4}{c}{Table A.2} \\
\hline
 field & R & $Z_2$ & $Z_{3H}$  \\
\hline
 $H_u$ &2 & 0& 0 \\
  $H_d$ &0 &0 &0  \\
 ${\cal U}_{1}$ & 0 &1 & 0  \\
 ${\cal U}_{2}$ & 0 &1 & 0 \\
 ${\cal U}_{3}$ & 0 &1 & 1 \\
 ${\cal D}_{1}$ & 1 &1 & 0 \\
 ${\cal D}_{2}$ & 1 &1 & 0 \\
 ${\cal D}_{3}$ & 0 &1 & 0 \\
\hline
\end{tabular}
\end{center}

Table A.2 gives an example of a model with a $Z_{3H}$ horizontal symmetry (configuration 2) in addition to matter parity. Model contains 1 up quark, 1 down quark and lepton suppressed by $\epsilon$, and two neutrinos.  All B and L violating operators are forbidden.

\begin{center}
\begin{tabular}{|c|c|c|c|}
\multicolumn{4}{c}{Table A.3} \\
\hline
 field & R & $Z_2$ & $Z_{4H}$  \\
\hline
 $H_u$ &2 & 0& 0 \\
  $H_d$ &0 &0 &0  \\
 ${\cal U}_{1}$ & 0 &1 & 2  \\
 ${\cal U}_{2}$ & 0 &1 & 1 \\
 ${\cal U}_{3}$ & 0 &1 & 3 \\
 ${\cal D}_{1}$ & 1 &1 & 0 \\
 ${\cal D}_{2}$ & 1 &1 & 0 \\
 ${\cal D}_{3}$ & 0 &1 & 1 \\
\hline
\end{tabular}
\end{center}

Table A.3 gives an example of a model with $Z_{4H}$ horizontal symmetry (configuration 2) in addition to matter parity.  Model contains heavy top, bottom and tau masses suppressed by $\epsilon$, charm mass suppressed by $\epsilon^2$, 2 neutrinos, and the rest of the quarks and leptons massless.  All B and L violating operators are forbidden.  The $Z_4$ R symmetry is anomaly free, however the $Z_{4H}$ symmetry is anomalous.

\begin{center}
\begin{tabular}{|c|ccc|cccccc|ccc|cccc|}
\multicolumn{17}{c}{Table A.4} \\
\hline
 config. & $H_u$ & $H_d$ & $p$ & ${\cal U}_{1}$  & ${\cal U}_{2}$ & ${\cal U}_{3}$ & ${\cal D}_{1}$ & ${\cal D}_{2}$ & ${\cal D}_{3}$ & B & L &  ${\cal U} {\cal U} {\cal U} {\cal D}$ & Ups & Downs & Lept. & Neut.   \\
\hline
 1st &2 &0 & 0& 0& 0& 0& 1& 1& 0& 16& 28& 1& 2& 1& 1& 2\\
 1st &2 &0 & 0& 0& 0& 0& 3& 3& 0& 16& 28& 1& 2& 1& 1& 2\\
 1st &2 &0 & 0& 0& 0& 0& 0& 1& 1& 16& 28& 1& 2& 1& 1& 2\\
 1st &2 &0 & 0& 0& 0& 0& 1& 0& 1& 16& 28& 1& 2& 1& 1& 2\\
 1st &2 &0 & 0& 0& 0& 0& 0& 3& 3& 16& 28& 1& 2& 1& 1& 2\\
 1st &2 &0 & 0& 0& 0& 0& 3& 0& 3& 16& 28& 1& 2& 1& 1& 2\\
 2nd &2 &0 & 1& 2& 2& 2& 1& 1& 2& 1& 7& 1& 1& 1& 0& 2\\
 2nd &2 &0 & 1& 2& 2& 2& 3& 3& 2& 1& 7& 1& 1& 1& 0& 2\\
\hline
\end{tabular}
\end{center}

Table A.4 shows anomaly-free models that have only one  ${\cal U} {\cal U} {\cal U} {\cal D}$ operator. These are good prospects for adding a discrete horizontal symmetry.

\begin{center}
\begin{tabular}{|c|ccc|cccccc|cc|cccc|}
\multicolumn{16}{c}{Table A.5} \\
\hline
 config & $H_u$ & $H_d$ & $p$ & ${\cal U}_{1}$  & ${\cal U}_{2}$ & ${\cal U}_{3}$ & ${\cal D}_{1}$ & ${\cal D}_{2}$ & ${\cal D}_{3}$ & B & L &  Ups & Downs & Lept. & Neut. \\
\hline
 1st & 0 & 2 & 0 & 2 & 2 &2 &1 &1 &1 &0&0 &1 &0 &0 &3 \\
 1st & 0 & 2 & 1 & 2 & 2 &2 &1 &1 &1 &0&0 &1 &0 &0 &3\\
 1st & 0 & 2 & 0 & 2 & 2 &2 &3 &3 &3 &0&0 &1 &0 &0 &3 \\
 1st & 0 & 2 & 1 & 2 & 2 &2 &3 &3 &3 &0&0 &1 &0 &0 &3 \\
\hline
\end{tabular}
\end{center}

Table A.5 shows 4 models with a massive top quark and 3 neutrinos which have no lepton or baryon-violating operators.  However, the top quark mass is suppressed by $\epsilon$.  These models are anomalous without the axion shift.

\begin{center}
\begin{tabular}{|c|ccc|cccccc|cc|cccc|}
\multicolumn{16}{c}{Table A.6} \\
\hline
 config & $H_u$ & $H_d$ & $p$ & ${\cal U}_{1}$  & ${\cal U}_{2}$ & ${\cal U}_{3}$ & ${\cal D}_{1}$ & ${\cal D}_{2}$ & ${\cal D}_{3}$ & B & L &  Ups & Downs & Lept. & Neut. \\
\hline
 2nd & 2 & 0 & 0 & 2 & 2 &2 &3 &3 &1 &12&0 &1 &0 &0 &3 \\
 2nd & 2 & 0 & 1 & 2 & 2 &2 &3 &3 &1 &12&0 &1 &0 &0 &3\\
 2nd & 2 & 0 & 0 & 2 & 2 &2 &1 &1 &3 &12&0 &1 &0 &0 &3 \\
 2nd & 2 & 0 & 1 & 2 & 2 &2 &1 &1 &3 &12&0 &1 &0 &0 &3 \\
\hline
\end{tabular}
\end{center}

Table A.6 shows 4 models (all anomalous) which have no lepton-violating operators. With an axion, these might make interesting models because they have some baryon-violation but no lepton violation, do not involve any extra discrete symmetries besides the R-symmetry, and could be phenomenologically acceptable if the R-symmetry is broken in just the right way.

\subsection{Models with a $10$ and $\overline{10}$}

\begin{center}
\begin{tabular}{|c|ccccc|cccccc|cc|cccc|}
\multicolumn{18}{c}{Table A.7} \\
\hline
config. & $H_u$ & $H_d$ & $p$ & $A$ & $B$ & ${\cal U}_{1}$  & ${\cal U}_{2}$ & ${\cal U}_{3}$ & ${\cal D}_{1}$ & ${\cal D}_{2}$ & ${\cal D}_{3}$ & B & L  & Ups & Downs & Lept. & Neut.   \\
\hline
  1st &0 &2 & 0 & 0 &2 &1 &1 &1 &1 &1 &1 &0 &0  &3 &3 &3 &3 \\
  1st  &0 &2 &0 & 0 &2 &3 &3 &3 &3 &3 &3 &0 &0  &3 &3 &3 &3 \\
1st &0 &2 &1 &2 &0 &1 &1 &1 &1 &1 &1 &0 &0  &3 &3 &3 &3 \\
  1st &0 &2 &1 & 2 &0 &3 &3 &3 &3 &3 &3 &0 &0  &3 &3 &3 &3 \\
\hline
\end{tabular}
\end{center}

Table A.7: Ideal models.  No B or L violation and full quark and lepton mass matrices.  All the up quarks are unsuppressed, downquarks and leptons are suppressed by $\epsilon$, and neutrinos are unsuppressed.  These models are ruled out by the instanton anomaly but are made possible by adding an axion.

\begin{center}
\begin{tabular}{|c|ccccc|cccccc|cc|cccc|}
\multicolumn{18}{c}{Table A.8} \\
\hline
config. & $H_u$ & $H_d$ & $p$ & $A$ & $B$ &  ${\cal U}_{1}$  & ${\cal U}_{2}$ & ${\cal U}_{3}$ & ${\cal D}_{1}$ & ${\cal D}_{2}$ & ${\cal D}_{3}$ & B  & L  & Ups & Downs & Lept. & Neut.   \\
\hline
 3rd& 2& 0& 0& 2& 0 &1 &1 &1 &1 &1 &3 &21 &21 &2 &3 &3 &3 \\
 3rd& 2& 0& 1& 0& 2 &1 &1 &1 &1 &1 &3 &21 &21 &2 &3 &3 &3 \\
 3rd& 2& 0& 0& 2& 0 &3 &3 &3 &3 &3 &1 &21 &21 &2 &3 &3 &3 \\
 3rd& 2& 0& 1& 0& 2 &3 &3 &3 &3 &3 &1 &21 &21 &2 &3 &3 &3 \\
\hline
\end{tabular}
\end{center}

Table A.8 shows models with everything massive (including neutrinos) except the up quark, anomaly-free, and a minimal number of ${\cal U} {\cal U} {\cal U} {\cal D}$
operators (21).  These occur in configuration 3.  There are none in configuration 2, and a lot in configuration 1 but all of those
have 27 ${\cal U} {\cal U} {\cal U} {\cal D}$  operators while these only have 21.  There are also a number of models that have only 8 ${\cal U} {\cal U} {\cal U} {\cal D}$ operators, but these have one neutrino massless.

\begin{center}
\begin{tabular}{|c|ccccc|cccccc|cc|cccc|}
\multicolumn{18}{c}{Table A.9} \\
\hline
config. & $H_u$ & $H_d$ & $p$ & $A$ & $B$ &  ${\cal U}_{1}$  & ${\cal U}_{2}$ & ${\cal U}_{3}$ & ${\cal D}_{1}$ & ${\cal D}_{2}$ & ${\cal D}_{3}$ & B  & L  & Ups & Downs & Lept. & Neut.   \\
\hline
 1st& 2& 0& 1& 0& 2 &0 &0 &0 &1 &1 &0 &15 &30 &3 &1 &1 &2 \\
 1st& 2& 0& 1& 0& 2 &0 &0 &0 &3 &3 &0 &15 &30 &3 &1 &1 &2 \\
 1st& 2& 0& 1& 0& 2 &0 &0 &0 &0 &1 &1 &15 &30 &3 &1 &1 &2 \\
 1st& 2& 0& 1& 0& 2 &0 &0 &0 &1 &0 &1 &15 &30 &3 &1 &1 &2 \\
 1st& 2& 0& 1& 0& 2 &0 &0 &0 &0 &3 &3 &15 &30 &3 &1 &1 &2 \\
 1st& 2& 0& 1& 0& 2 &0 &0 &0 &3 &0 &3 &15 &30 &3 &1 &1 &2 \\
\hline
 2nd& 0& 2& 0& 0& 2 &0 &0 &0 &1 &1 &0 &15 &24 &1 &1 &1 &2 \\
 2nd& 0& 2& 0& 0& 2 &0 &0 &0 &3 &3 &0 &15 &24 &1 &1 &1 &2 \\
 2nd& 1& 1& 1& 3& 3 &1 &1 &1 &0 &0 &1 &12 &6  &0 &1 &0 &2 \\
 2nd& 1& 1& 1& 3& 3 &1 &1 &1 &2 &2 &1 &18 &42 &0 &2 &2 &2 \\
 2nd& 2& 0& 0& 2& 0 &2 &2 &2 &1 &1 &2 &0  &9  &2 &1 &0 &2 \\
 2nd& 2& 0& 0& 2& 0 &2 &2 &2 &3 &3 &2 &0  &9  &2 &1 &0 &2 \\
 2nd& 3& 3& 1& 1& 1 &3 &3 &3 &0 &0 &3 &12 &6  &0 &1 &0 &2 \\
 2nd& 3& 3& 1& 1& 1 &3 &3 &3 &2 &2 &3 &18 &42 &0 &2 &2 &2 \\
\hline
 3rd& 2& 0& 0& 0& 2 &0 &0 &0 &1 &1 &0 &15 &24 &1 &1 &1 &2 \\
 3rd& 2& 0& 0& 0& 2 &0 &0 &0 &3 &3 &0 &15 &24 &1 &1 &1 &2 \\
 3rd& 3& 3& 1& 3& 3 &1 &1 &1 &0 &0 &1 &12 &6  &0 &1 &0 &2 \\
 3rd& 3& 3& 1& 3& 3 &1 &1 &1 &2 &2 &1 &18 &42 &0 &2 &2 &2 \\
 3rd& 0& 2& 0& 2& 0 &2 &2 &2 &1 &1 &2 &0  &9  &2 &1 &0 &2 \\
 3rd& 0& 2& 0& 2& 0 &2 &2 &2 &3 &3 &2 &0  &9  &2 &1 &0 &2 \\
 3rd& 1& 1& 1& 1& 1 &3 &3 &3 &0 &0 &3 &12 &6  &0 &1 &0 &2 \\
 3rd& 1& 1& 1& 1& 1 &3 &3 &3 &2 &2 &3 &18 &42 &0 &2 &2 &2 \\
\hline
\end{tabular}
\end{center}

Table A.9 shows anomaly-free models from all three configurations that have non-trivial mass matrices and no ${\cal UUUD}$ operators.  The remaining B and L violating operators can be forbidden by an anomaly-free $Z_2$ matter parity.

\begin{center}
\begin{tabular}{|c|ccccc|cccccc|cc|cccc|}
\multicolumn{18}{c}{Table A.10} \\
\hline
 config. & $H_u$ & $H_d$ & $A$ & $B$ & $p$ & ${\cal U}_{1}$  & ${\cal U}_{2}$ & ${\cal U}_{3}$ & ${\cal D}_{1}$ & ${\cal D}_{2}$ & ${\cal D}_{3}$ & B  & L &  Ups & Downs & Lept. & Neut.   \\
\hline
1st  & 3& 3&3 &3 &0 &2 &2 &2 &1 &2 &1 &0 &3   &0 & 0&0&1 \\
1st  & 0& 2&2 &0 &1 &2 &2 &2 &1 &2 &1 &0 &3   &0 & 0&0&2 \\
1st  & 3& 3&1 &1 &1 &2 &2 &2 &1 &2 &1 &0 &3   &0 & 0&0&1 \\
1st  & 3& 3&3 &3 &0 &2 &2 &2 &2 &1 &1 &0 &3   &0 & 0&0&1 \\
1st  & 0& 2&2 &0 &1 &2 &2 &2 &2 &1 &1 &0 &3   &0 & 0&0&2 \\
1st  & 3& 3&1 &1 &1 &2 &2 &2 &2 &1 &1 &0 &3   &0 & 0&0&1 \\
1st  & 3& 3&3 &3 &0 &2 &2 &2 &1 &1 &2 &0 &3   &0 & 0&0&0 \\
1st  & 0& 2&2 &0 &1 &2 &2 &2 &1 &1 &2 &0 &3   &0 & 0&0&2 \\
1st  & 3& 3&1 &1 &1 &2 &2 &2 &1 &1 &2 &0 &3   &0 & 0&0&0 \\
1st  & 1& 1&1 &1 &0 &2 &2 &2 &3 &3 &2 &0 &3   &0 & 0&0&0 \\
1st  & 0& 2&2 &0 &1 &2 &2 &2 &3 &3 &2 &0 &3   &0 & 0&0&2 \\
1st  & 1& 1&3 &3 &1 &2 &2 &2 &3 &3 &2 &0 &3   &0 & 0&0&0 \\
1st  & 1& 1&1 &1 &0 &2 &2 &2 &2 &3 &3 &0 &3   &0 & 0&0&1 \\
1st  & 0& 2&2 &0 &1 &2 &2 &2 &2 &3 &3 &0 &3   &0 & 0&0&2 \\
1st  & 1& 1&3 &3 &1 &2 &2 &2 &2 &3 &3 &0 &3   &0 & 0&0&1 \\
1st  & 1& 1&1 &1 &0 &2 &2 &2 &3 &2 &3 &0 &3   &0 & 0&0&1 \\
1st  & 0& 2&2 &0 &1 &2 &2 &2 &3 &2 &3 &0 &3   &0 & 0&0&2 \\
1st  & 1& 1&3 &3 &1 &2 &2 &2 &3 &2 &3 &0 &3   &0 & 0&0&1 \\
\hline
2nd  & 2& 0&2 &0 &0 &2 &2 &2 &1 &1 &2 &0 &9   &2 & 1&0&2 \\
2nd  & 3& 3&3 &3 &0 &2 &2 &2 &1 &1 &2 &0 &9   &0 & 1&0&1 \\
2nd  & 3& 3&1 &1 &1 &2 &2 &2 &1 &1 &2 &0 &9   &0 & 1&0&1\\
2nd  & 1& 1&1 &1 &0 &2 &2 &2 &3 &3 &2 &0 &9   &0 & 1&0&1 \\
2nd  & 2& 0&2 &0 &0 &2 &2 &2 &3 &3 &2 &0 &9   &2 & 1&0&2 \\
2nd  & 1& 1&3 &3 &1 &2 &2 &2 &3 &3 &2 &0 &9   &0 & 1& 0&1 \\
\hline
3rd  & 0& 2&2 &0 &0 &2 &2 &2 &1 &1 &2 &0 &9   &2 & 1&0&2 \\
3rd  & 1& 1&3 &3 &0 &2 &2 &2 &1 &1 &2 &0 &9   &0 & 1&0&1 \\
3rd  & 1& 1&1 &1 &1 &2 &2 &2 &1 &1 &2 &0 &9   &0 & 1&0&1\\
3rd  & 0& 2&2 &0 &0 &2 &2 &2 &3 &3 &2 &0 &9   &2 & 1&0&2 \\
3rd  & 3& 3&1 &1 &0 &2 &2 &2 &3 &3 &2 &0 &9   &0 & 1&0&1 \\
3rd  & 3& 3&3 &3 &1 &2 &2 &2 &3 &3 &2 &0 &9   &0 & 1& 0&1 \\
\hline
\end{tabular}
\end{center}

Table A.10 shows anomaly-free models with no B violating operators

\end{changemargin}

%
%
%
%
%
%

\end{document}